\documentclass[journal]{IEEEtran}

\usepackage{graphicx}
\usepackage{amssymb}
\usepackage{epstopdf}
\usepackage{url}

\usepackage{amsmath}
\usepackage{amsfonts}
\usepackage{mathtools}
\usepackage{booktabs}
\usepackage{threeparttable}
\usepackage{float}
\usepackage{color,soul}

\usepackage{booktabs}
\usepackage{tabularx}
\usepackage{tabu}
\usepackage{multirow}

\ifCLASSINFOpdf
\else
\fi
\begin{document}
%
\title{Secure and Efficient Multi-Signature Schemes for Fabric: An Enterprise Blockchain Platform}
%
%
%

\author{Yue Xiao, Peng Zhang, Yuhong Liu
\thanks{Peng Zhang is the corresponding author.}
\thanks{This work was in part supported by the National Natural Science Foundation of China (61702342, 61872243).}
\thanks{Y. Xiao and P. Zhang are with the College
of Electronics and Information Engineering, Shenzhen University, Shenzhen 518060, China (e-mail: xiaoyue2017@email.szu.edu.cn; zhangp@szu.edu.cn ).}
\thanks{Y. Liu is with the Department of Computer Engineering, Santa Clara University, Santa Clara 95053, USA (e-mail: yhliu@scu.edu).}
}

%
%

\markboth{Journal of \LaTeX\ Class Files}
{Shell \MakeLowercase{\textit{et al.}}: Bare Demo of IEEEtran.cls for IEEE Journals}
%



\maketitle

\begin{abstract}
Digital signature is a major component of transactions on Blockchain platforms, especially in enterprise Blockchain platforms, where multiple signatures from a set of peers need to be produced to endorse a transaction. However, such process is often complex and time-consuming. Multi-signature, which can improve transaction efficiency by having a set of signers cooperate to produce a joint signature, has attracted extensive attentions. In this work, we propose two multi-signature schemes, GMS and AGMS, which are proved to be more secure and efficient than state-of-the-art multi-signature schemes. Besides, we implement the proposed schemes in a real Enterprise Blockchain platform, Fabric. Experiment results show that the proposed AGMS scheme helps achieve the goal of high transaction efficiency, low storage complexity, as well as high robustness against rogue-key attacks and \(k\)-sum problem attacks.
\end{abstract}


%
\IEEEpeerreviewmaketitle

\section{Introduction}
%
%
%
%
\IEEEPARstart{A}{s} an emerging distributed ledger technology, Blockchain \cite{08nakamoto2008bitcoin} has shown great potential to transform business and finance fields. Recently, several banks, such as J.P. Morgan and Banco Santander S.A., have started to launch Blockchain based platforms in capital markets, which are characterized by ``huge sums of money, multiple stakeholders and lots of coordination" \cite{JP_Morgan}. As transactions in capital markets often require approvals from multiple parties, where each party has to identify whether information matches transaction history and follows the rules created by the participants, the approval process is often complex and time consuming. It is believed that Blockchain can effectively help cut costs and smooth transactions among multiple parties \cite{JP_Morgan}.

It is worth mentioning that Fabric \cite{11DBLP:conf/eurosys/AndroulakiBBCCC18}, an open-source permissioned Blockchain platform for enterprise use cases, has enabled endorsement functions to allow a set of endorsers to approve the execution of a transaction. Cryptographic digital signatures have been adopted to guarantee the validity of endorsements from all endorsers before a transaction can be added to the Blockchain ledger.

However, the endorsement process based on cryptographic digital signatures is often resource consuming, inefficient, and lack of scalability. In particular, to avoid inconsistency in transaction states, a signature needs to be collected from each endorser according to the endorsement policy. The verification of these signatures consumes large amounts of computational resources. After verification, these signatures, which can occupy a significant amount of storage space in a transaction, will be stored in a block and broadcast over the entire Blockchain network. Due to the large computation and communication overhead, the overall throughput of Fabric is about 100 to 2000 tps, which is very low and easily leads to network transmission delay.

A promising approach to improve the throughput is multi-signature \cite{18itakura1983public}, which allows a group of users to sign on a single message, and produces a joint signature that stands for all signers' agreement on the message. Generally, a joint signature has the same length as a single signature, and only needs to be verified once with the public keys of signers that participate. Therefore, compared to digital signature, multi-signature has many advantages such as lower bandwidth, less storage space, and faster verification. Multi-signature has been applied in many fields, including distributed certificate authorities \cite{45DBLP:conf/ccs/SzalachowskiMP14}, directory authorities \cite{38DBLP:journals/corr/SytaTVWF15}, and timestamping services \cite{02DBLP:conf/sp/SytaTVWJGGKF16}.

There are three major categories of multi-signature schemes, as RSA-based, BLS-based, and Schnorr-based multi-signature schemes. Compared to the other two types of schemes, the Schnorr-based multi-signature schemes can well balance the trade off between computational complexity and required storage space, and therefore attract extensive research attentions recently. For example, based on Schnorr signature \cite{09DBLP:journals/joc/Schnorr91}, BN multi-signature scheme \cite{04DBLP:conf/ccs/BellareN06} is designed by adding one more round in signing algorithm. BCJ multi-signature scheme \cite{35DBLP:conf/ccs/BagherzandiCJ08} is presented to eliminate the adding round by homomorphic trapdoor commitments. Gregory et al. design Musig multi-signature scheme \cite{DBLP:journals/dcc/MaxwellPSW19} to improve BN scheme. One of the most popular multi-signature schemes is CoSi \cite{02DBLP:conf/sp/SytaTVWJGGKF16}, which introduces a spanning tree structure to make it easily scale to thousands of signers. However, CoSi can be easily forged by rogue-key attacks and \(k\)-sum problem attacks \cite{03drijverssecurity}. Also, the leader with excessive power in CoSi may replace the message \(m\) to produce another challenge \(c^\prime\).

In this work, we aim to fill the research gap by proposing secure and efficient multi-signature schemes, which can decrease the storage of each transaction, improve the transmission rate of block, and shorten the verification and update time of each node. Our major contributions are described as follows.

\begin{itemize}
    \item Based on Gamma signature \cite{01DBLP:journals/tifs/YaoZ13}, we propose a secure multi-signature scheme named GMS (Gamma Multi-Signature) using proof of possession, which is robust against rogue-key attacks and \(k\)-sum problem attacks. It also addresses the problem of excessive power of the leader in CoSi. In addition, the proposed GMS has achieved strong provable security.

    \item To further improve the online performance of GMS, we propose the Advanced Gamma Multi-Signature (AGMS), a more efficient multi-signature scheme. In particular, we propose to change the running order of phases in the signing algorithm to reduce calculation steps after message arrivals. In addition, by enabling the key aggregation algorithm to run together with the signing algorithm, the distributed execution of the key aggregation algorithm is allowed, which further reduces the overall execution time.

    \item Based on the proposed AGMS scheme, we improve the transaction process in Fabric, for which we deploy the multi-signature in and aggregate multiple signatures from endorsers to a joint signature, so as to reduce the size of the transaction and improve the efficiency of endorsement and ledger update. The implementation results also show that our designed transaction process can successfully improve the efficiency and throughput of Fabric.
\end{itemize}

The rest of this paper is organized as follows. Related works are summarized in Section II, followed by preliminaries in Section III. In Section IV, we discuss the two proposed multi-signature schemes GMS and AGMS. The corresponding security analysis and performance analysis are presented in Section V and VI respectively. Finally, the application to Fabric is described in Section VII and Section VIII provides the conclusion.

\vspace{-2mm}
\section{Related work}
According to the difficulty assumptions and basic signature algorithms, multi-signature schemes can be divided into RSA-based, BLS-based, Schnorr-based, etc. The details are described as follows.
\subsection{Multi-signature schemes derived from RSA signatures}
As the implementation of RSA is particularly efficient, there are some multi-signature schemes proposed under RSA assumption.
Harn et al. \cite{1989New} propose a multi-signature scheme based on RSA for the first time, for which the time to generate and verify multiple signatures depends on the number of signers.
Bellare and Neven \cite{DBLP:conf/ctrsa/BellareN07} propose an identity-based multi-signature scheme which relies on the RSA assumption in the random oracle model. The scheme has fast multi-signature generation and verification, but it takes three rounds of interactions.
Based on \cite{DBLP:conf/ctrsa/BellareN07}, Bagherzandi et al. \cite{DBLP:conf/pkc/BagherzandiJ10} propose an improved identity-based multi-signature scheme and aggregation signature scheme under RSA assumptions. The number of interactive rounds of the scheme is reduced from three to two.
Tsai et al. \cite{DBLP:conf/isbast/TsaiLW13} propose an identity-based sequential aggregation signature scheme which can be seen as a generalization of multi-signature, where each signer signs a different message, and signatures are aggregated in sequence.
Hohenberger et al. \cite{DBLP:conf/eurocrypt/HohenbergerW18} construct a synchronized aggregation signature from RSA, which can be used in Blockchain so that the creation of a new block can be seen as a synchronization event.
Yu et al. \cite{44DBLP:journals/ijdsn/YuZWGXD0Y18} propose the use of multi-signature and Blockchain to ensure security and privacy of the transmitted data in the Internet of Things (IoT) scenario.
Compared to the schemes derived from Schnorr signature, length of signatures in these RSA based schemes is significantly longer for a similar level of security.

\subsection{Multi-signature schemes derived from BLS signatures}

BLS signature \cite{DBLP:conf/asiacrypt/BonehLS01} is proposed based on bilinear paring, where the signature length is just 224-bit compared to the 2048-bit signature in RSA. Based on efficient bilinear parings and elegant BLS signatures, various multi-signature schemes \cite{DBLP:conf/pkc/Boldyreva03}\cite{DBLP:conf/eurocrypt/RistenpartY07}\cite{DBLP:conf/ccs/AmbrosinCINSS16}\cite{07DBLP:conf/asiacrypt/BonehDN18} are proposed.
Particularly, Ambrosin et al. \cite{DBLP:conf/ccs/AmbrosinCINSS16} propose a novel optimistic aggregation signature scheme called OAS to design secure collective attestation for Internet of Things.
Boneh et al. \cite{07DBLP:conf/asiacrypt/BonehDN18} also propose a BLS multi-signatures with public-key aggregation in order to reduce the size of Bitcoin Blockchain.
Compared to the schemes derived from Schnorr signature, these bilinear pairing based schemes can further reduce the key and signature sizes. However, as the bilinear pairing operation is one of the most complex operations in modern cryptography \cite{DBLP:journals/tifs/HeZXH15}, they also introduce high computational overhead.

\subsection{Multi-signature schemes derived from Schnorr signatures}
When one uses a 2048-bit modulus, the corresponding signature lengths for RSA, BLS, and Schnorr based schemes are 2048 bits, 224 bits, and 448 bits, respectively. Although the advantage of BLS signature length is obvious, the high computational cost can not be ignored. Considering both computation and storage, Schnorr signature \cite{09DBLP:journals/joc/Schnorr91}, one of the best-known signature algorithms, is a good choice.
Many multi-signature schemes are proposed based on Schnorr signature.
Bellare and Neven \cite{04DBLP:conf/ccs/BellareN06} have designed BN scheme by adding one more round in the signing algorithm, where all signers involved need to exchange their own commitments. It is proved secure in the plain public-key model.
Then, Bagherzandi et al. \cite{35DBLP:conf/ccs/BagherzandiCJ08} propose BCJ scheme to eliminate the adding round by using homomorphic trapdoor commitments.
Gregory et al. \cite{DBLP:journals/dcc/MaxwellPSW19} design Musig scheme to improve BN scheme in two aspects: holding the same key and signature size with Schnorr signature, and allowing key aggregation. Furthermore, Musig scheme is also applied to Bitcoin network to support key aggregation without revealing the individual signer's public key.

One of the most popular Schnorr based multi-signature schemes is CoSi \cite{02DBLP:conf/sp/SytaTVWJGGKF16}, which requires each node to sign the same message \(m\) by communicating and computing bottom-up in a spanning tree structure. The introduction of the spanning tree structure makes it easy for CoSi to scale up to thousands of signers. Because of its great scalability, CoSi has served as a basis for many multi-signature schemes proposed in later research works \cite{31alangot2018reliable}\cite{28DBLP:conf/uss/Kokoris-KogiasJ16}\cite{29DBLP:conf/sp/SytaJKGGKFF17}\cite{30DBLP:conf/trustcom/ZhouWQHL16}. However, Drijvers et al. \cite{03drijverssecurity} point out that CoSi can be easily forged by rogue-key attacks and \(k\)-sum problem attacks. The leader in CoSi can also forge a joint signature on another message \(m^\prime\) without any other information. Therefore, mBCJ, a new multi-signature scheme modified from CoSi, is proposed to defend against these attacks. Nevertheless, the computation of mBCJ is more complicated and time-consuming. As a summary, although the security of CoSi is challenged, it is efficient and scalable. Although proof of possession can be introduced to improve the security, it will potentially increase the overall computational costs.

Therefore, considering both security and efficiency, we turn to consider other digital signature schemes. Gamma signature \cite{01DBLP:journals/tifs/YaoZ13}, proposed by Yao et al. in 2013, is modified from Schnorr signature. Different from Schnorr signature, Gamma signature can be implemented in two corresponding phases: the offline phase, which pre-computes some partial values without any information of the message \(m\) to be signed, and the online phase, which produces the final signature after the message \(m\) arrives. Compared to Schnorr signature, Gamma signature performs better in several aspects: (1) online performance; (2) flexible and easy deployment in interactive protocols; and (3) great unforgeability against concurrent interactive attacks. To our best knowledge, this work is the first multi-signature scheme based on Gamma signature. Experiment results verify that better online performance can be achieved when compared to the above mentioned Schnorr-based multi-signature schemes.


\vspace{-2mm}
\section{Preliminaries}
\subsection{Target One-Way Hash Function}

\noindent\emph{\bf Definition 1 (Target One-Way Hash Function \cite{01DBLP:journals/tifs/YaoZ13}):}
A hash function \(H:\{0,1\}^*\rightarrow\varepsilon\subseteq{\{0,1\}}^{l_0}\) is defined as a \((t_f,\varepsilon_f)\) target one-way hash function w.r.t. an \(e\)-condition \(R_e\) and a set \(D\subseteq\{0,1\}^{l_0}\), if for any probabilistic poly-time adversary \(\mathcal{A}\), there exists a relationship that
\begin{displaymath}
\begin{array}{l}
\textup{Adv}^{\textsf{tow}}_{H}(\mathcal{A})=\\\\
\textup{Pr}
\left[
\begin{array}{l}
R_e(d,e,d^\prime,e^\prime)=0
\end{array}
\middle|
\begin{array}{l}
(m,s)\leftarrow A_1(H,d)\\
m^\prime\leftarrow A_2(H,d,m,d^\prime,s)
\end{array}
\right]\\\\
\leq\textsf{negl} (l_0),
\end{array}
\end{displaymath}
where for any \(t\)-time algorithm \(A=\{A_1,A_2\}\), we assume that  \(e=H(m)\), \(e^\prime=H(m^\prime)\), \(d,d^\prime\leftarrow D\), and \(s\) is defined as some state information passed from \(A_1\) to \(A_2\).

\subsection{Rogue-Key Attack and \(k\)-Sum Problem Attack}

Rogue-key attack is a very typical attack against multi-signature schemes including CoSi and BN, allowing a corrupted signer to set his/her own public key arbitrarily such as \(X_1=g_1^{sk_1}(\prod_{i=2}^{n} X_i)^{-1}\) so that he/she can independently forge a joint signature on the same messages \(m\) for public keys \(\{X_1,...,X_n\}\).

To protect systems against rogue-key attacks, some researchers choose to use a sophisticated key generation protocol. For example, proof of possession, proposed by Ristenpart and Yilek \cite{05DBLP:conf/eurocrypt/RistenpartY07}, is a direct way to defend against this attack.
It is established based on the general key registered model, meaning that the signer is required to provide his/her knowledge of the secret key \(sk\) corresponding to the public key \(pk\) through a non-interactive zero knowledge protocol. This proof is able to stop the corrupted signer forging a joint signature. It is suitable to be applied in Public Key Infrastructure (PKI), where each node has its certificate showing the information about its own public key \(pk\).

In addition, as stated in \cite{03drijverssecurity}, there exists \(k\)-sum problem attack that belongs to a \(k\)-dimensional generalization of the birthday problem. It can effectively compromise several multi-signature schemes, such as CoSi \cite{02DBLP:conf/sp/SytaTVWJGGKF16}, Musig \cite{DBLP:journals/dcc/MaxwellPSW19}. In particular, the \(k\)-sum problem is defined as follows.

\noindent\emph{\bf Definition 2 (\(k\)-Sum Problem \cite{03drijverssecurity}):} Given a group \((\mathbb{Z}_q,+)\), an arbitrary \(l_0\)-bit prime \(q\), and \(k\) lists  \(L_1,\cdots,L_k\) with an identical size, where elements in each list are sampled uniformly and randomly from \(\mathbb{Z}_q\), the \(k\)-sum problem aims to find out \(k\) values: \(x_1\in L_1, \cdots,x_k \in L_k\) that satisfy the equation \(x_1+\cdots+x_k\equiv0\ mod\ q\).

We consider that an adversary can successfully launch a \(k\)-sum problem attack if he/she can solve the \(k\)-sum problem by using \(k\) lists with length of \(s_L\), within a total running time of \(\tau\) and certain probability that


\begin{displaymath}
\begin{array}{l}
\textup{Adv}^{\textsf{\(k\)-sum}}_{\mathbb{Z}_q}({\mathcal{A}})=\\\\
\textup{Pr}
\left[
\begin{array}{l}
x_1+\cdots+x_k\equiv0\ mod\ q
\end{array}
\middle|
\begin{array}{l}
L_1,\cdots,L_k\in\mathbb{Z}_q\\
|L_1|=\cdots =|L_k|=s_L\\
x_1\in L_1,\cdots ,x_k\in L_k
\end{array}
\right]\\\\
\geq \textsf{negl} (l_0).
\end{array}
\end{displaymath}

According to the above construction of \(k\)-sum problem, the adversary working as a leader in CoSi needs to simulate the signing algorithm \((k-1)\) times to produce different joint signatures on the same message \(m\), so that it can forge a joint signature on a new message \(m^\prime\) satisfying \(k\)-sum problem. Therefore, an effective way to avoid this attack is to improve the construction of the signing algorithm.

However, rogue-key attack and \(k\)-sum problem attack are not handled in CoSi. Therefore, we propose to adopt proof of possession in key generation algorithm and improve signing algorithm, so that secure multi-signature schemes can be developed against rogue-key attacks and \(k\)-sum problem attacks.

\subsection{Gamma Signature}
The improvements to resist attacks and guarantee security will inevitably increase the total computational costs. Hence it is very challenging to consider security and efficiency at the same time. Nevertheless, if we are able to move part of the computational overhead from online to offline, an improvement on both security and online efficiency may be achieved even if the total computational costs (i.e. including both online and offline) are higher.

Gamma signature \cite{01DBLP:journals/tifs/YaoZ13} is such an online/offline signature scheme, which has better online performance compared to Schnorr signature. In particular, it is implemented in two corresponding phases: the offline phase, which pre-computes some partial values without any information of the message \(m\) to be signed, and the online phase, which produces the final signature after the arrival of message \(m\). The detailed procedure of Gamma signature is explained as follows.\\

%

\noindent\textbf{Parameter generation.} We use \(\textsf{Pg}(\kappa)\) to set up a group \(\mathbb{G}\) of order \(q\)  with generator \(g_1\), where \(q\) is defined as a prime with  \(\kappa\)-bit, and finally output \(par=(\mathbb{G},g_1,q)\).

\noindent\textbf{Key generation.} \(\textsf{Kg}(par)\) randomly selects \(sk\in [0,q-1]\), computes \(pk= g_1^{sk}\) and finally outputs value \((pk,sk)\).

\noindent\textbf{Signing.} This algorithm defines two kinds of hash functions: \(H_0:\{0,1\}^*\rightarrow\mathbb{Z}_q\) that is modelled as random oracles and \(H_1:\{0,1\}^*\rightarrow\mathbb{Z}_q\) that belongs to a target one-way hash function. A signer runs \(\textsf{Sign}(par,sk,m)\) by first randomly selecting a value \(v\in[0,q-1]\) and pre-computing \(V=g_1^v\ mod\ q\), \(c=H_0(V,pk)\), and \(v*c\). When the message \(m\) comes, the signer can further compute \(e=H_1(m)\) and \(s=v*c-e*sk\ mod\ q\), and output \(\sigma= (c,s)\) as a signature on the message \(m\).

\noindent\textbf{Verification.} To run \(\textsf{Vf}(par,pk,m,\sigma)\), the verifier firstly computes \(e=H_1(m)\), \(V=(g_1^s*pk^e)^{c^{-1}}\ mod\ q\), and then checks whether it satisfies \(H_0(V,pk)=c\). If not, the signature is invalid and the verifier rejects it. Else, the verifier accepts the signature.

Due to its high online efficiency, Gamma signature is adopted in this paper as a basis for the proposed multi-signature schemes. To our best knowledge, this is the first work that proposes multi-signature schemes based on Gamma signature.

\section{Proposed Multi-Signature Schemes}
As mentioned before, CoSi is an efficient and scalable multi-signature scheme, but it is easily forged by rogue-key attacks and \(k\)-sum problem attacks. The leader in CoSi can also forge a joint signature by producing the final challenge \(c^\prime\) on another message \(m^\prime\). It is of great significance to design a new multi-signature scheme with enhanced security, high scalability, and efficiency.

\subsection{Gamma Multi-Signature Scheme}
With the motivation of constructing a more secure, efficient, and scalable multi-signature scheme, we propose a new multi-signature scheme. In particular, we introduce proof of possession to ensure security of the proposed scheme against rogue-key attacks. To reduce the extra computational costs introduced by proof of possession, we adopt Gamma signature \cite{01DBLP:journals/tifs/YaoZ13} as the basis to split the overall computation into online and offline parts, so that the computational complexity for the online part is improved when compared to CoSi signature scheme and make it hard to forge by \(k\)-sum problem attacks. Furthermore, inspired by CoSi, we also adopt the spanning tree structure to improve the scalability of the proposed scheme. As a summary, our design goal is to ensure security against the rogue-key attacks and \(k\)-sum problem attacks, while achieving high online efficiency and scalability.


We firstly propose Gamma Multi-Signature (GMS) scheme. Assume our proposed multi-signature scheme GMS consists of six algorithms \(\textsf{GMS}=\{\textsf{Pg},\textsf{Kg},\textsf{KAg},\textsf{Sign},\textsf{KVf},\textsf{Vf}\}\) and adopts four hash functions:  \(H_0,H_1,H_2,H_3:\{0,1\}^*\rightarrow\mathbb{Z}_q\), where \(H_0\), \(H_1\) are modelled as random oracles and \(H_2\), \(H_3\) are target one-way hash functions. It works as follows.\\

\noindent\textbf{Parameter generation.} We use \(\textsf{Pg}(\kappa)\) to set up a group \(\mathbb{G}\) of order \(q\)  with generator \(g_1\), where \(q\) is defined as a prime with  \(\kappa\)-bit, and finally outputs \(par=(\mathbb{G},g_1,q)\).

\noindent\textbf{Key generation.} \(\textsf{Kg}(par)\) randomly picks \(sk\in [0,q-1]\) as a private key and sets \(y = g_1^{sk}\) as the corresponding public key. Then, it constructs proof of possession \(\pi=(a,d)\) of \(sk\), which is to protect against rogue-key attacks, by choosing \(r\stackrel{\$}{\leftarrow} \mathbb{Z}_q\) and computing \(a= H_1(g_1,g_1^r)\), \(b= H_2(y)\), and \(d= r*a-b*sk\ mod\ q\). Finally, it sets \(pk= (y,\pi)\) and outputs \((pk,sk)\). The proof of possession will be checked by the verifier each time when a new key pair involved to sign is found. Proof of possession is used to defend against rogue-key attacks existing in CoSi.

\noindent\textbf{Key Aggregation.} Given \(\mathcal{PK}\) as the set of all public keys, \(\textsf{KAg}(\mathcal{PK})\) parses each public key \({pk_i}\) involved to sign in \(\mathcal{PK}\) as \( pk_i=(y_i,\pi_i)\),
and outputs the aggregated public key as \(\tilde{X}= \prod_{pk_i\in\mathcal{PK}}{y_i}\).

\begin{figure}
\includegraphics[width=0.5\textwidth]{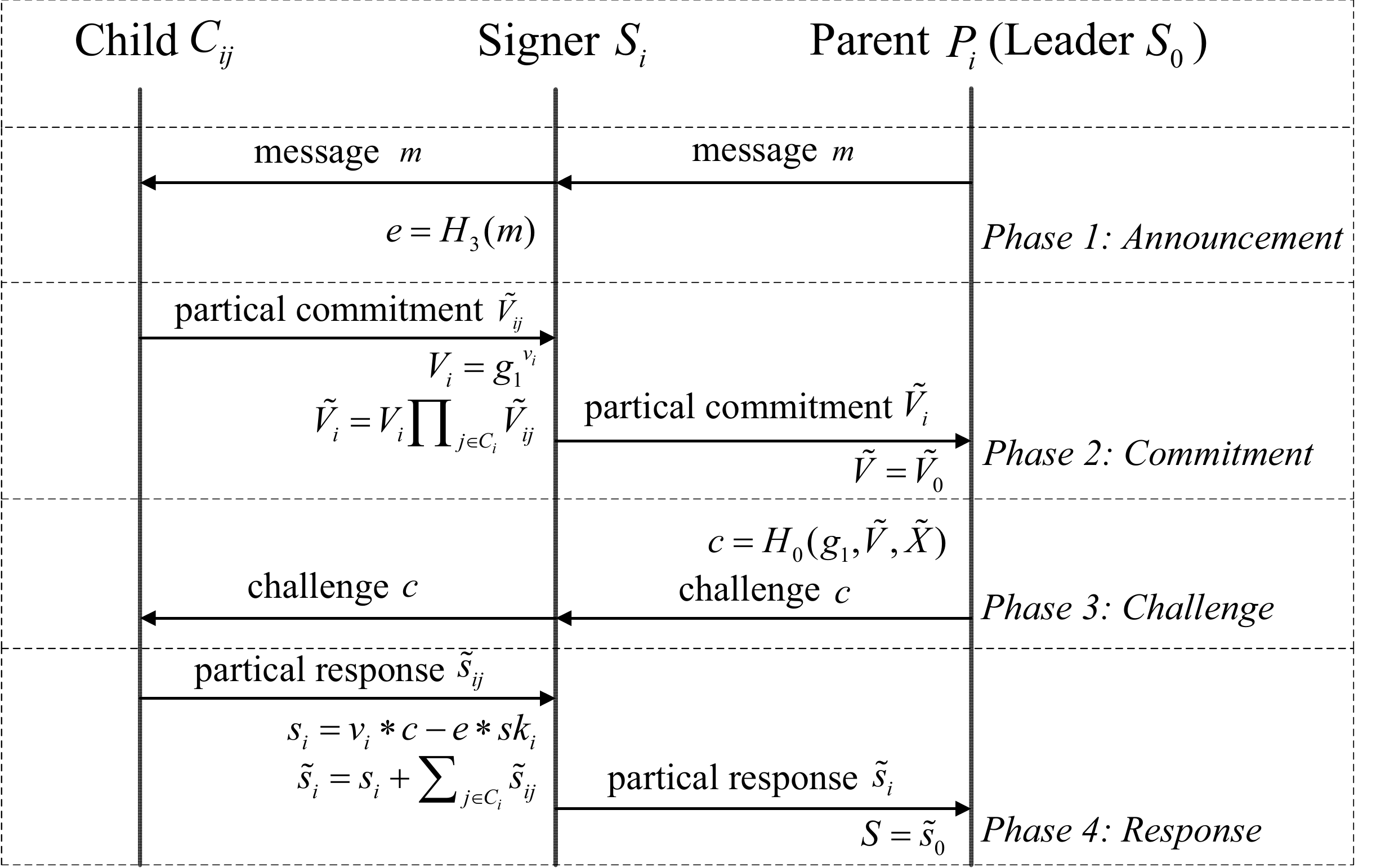}
\caption{The signing algorithm of our GMS scheme (We suppose that signer \(S_i\) holds the key pair \((pk_i,sk_i)\), where \(pk_i=(y_i,\pi_i)\), and parent \(P_i\) works as a leader \(S_0\). If parent \(P_i\) is not a leader, it just works as signer \(S_i\). Finally, the leader \(S_0\) outputs \((c,S)\) as the joint signature.)}
\label{figGMS}
\end{figure}

\noindent\textbf{Signing.} We set \(C_i=\{C_{ij}\}\) as the set of children of one signer \(S_i\) in the spanning tree structure, and \(P_i\) as the parent of signer \(S_i\). Assume \(S_0\) to be the root of the tree, so called the leader. The signer \(S_i\) runs signing algorithm \(\textsf{Sign}(par,sk_i,m,\tau)\) in a tree \(\tau\) for four phases, which is shown in Fig. \ref{figGMS}.

\noindent\emph{\underline{Phase 1: Announcement.}} When the leader \(S_0\) receives a message \(m\), it starts to multicast the announcement \(m\) to its children top-down in the tree structure.

\noindent\emph{\underline{Phase 2: Commitment.}} This process is run in a bottom-up way by each node \(S_i\). Specifically, given a node \(S_i\), after receiving the announcement \(m\), \(S_i\) firstly chooses a random secret value \(v_i\) and computes \(V_i=g_1^{v_i}\). Then, \(S_i\) waits for each immediate child \(j\)'s partial commitment \(\tilde{V}_{ij}\). When all the partial commitments are received, \(S_i\) computes \(\tilde{V}_i=V_i\prod_{j\in C_i} \tilde{V}_{ij}\). After that, the result \(\tilde{V}_i\) is send to its parent \(P_i\) unless \(S_i\) is the leader (i.e. \(i=0\)).

\noindent\emph{\underline{Phase 3: Challenge.}} The leader \(S_0\) waits for each immediate child's partial commitment value \(\tilde{V}_{0j}\) and computes the final commitment \(\tilde{V}=\tilde{V}_0=V_0\prod_{j\in C_0} \tilde{V}_{0j}\). So, the collective challenge is \(c=  H_0(g_1,\tilde{V},\tilde{X})\). The value \(c\), as a part of the joint signature, can be sent to the verifier in advance or stored at the leader. After that, the leader sends the shared challenge value \(c\) back to its children.

\noindent\emph{\underline{Phase 4: Response.}} When \(S_i\) receives \(c\), it can compute the response: \(s_i=v_i*c-e*sk_i\), where \(c= H_0(g_1,\tilde{V},\tilde{X})\) and \(e=H_3(m)\), and wait for each partial response \(\tilde{s}_{ij}\) from its immediate children \(j\). When all the partial responses are received, it sets \(\tilde{s}_i= s_i+\sum_{j\in C_i}{\tilde{s}_{ij}}\). After that, the result \(\tilde{s}_i\) is sent to its parent \(P_i\) unless \(S_i\) is the leader (i.e. \(i=0\)). Finally, the leader \(S_0\) computes the final response \(S= \tilde{s}_0=s_0+\sum_{j\in C_0}{\tilde{s}_{0j}}\) and outputs the joint signature \((c,S)\).

Compared to CoSi, we divide the challenge \(c\) into two independent values \(c\) and \(e\), so as to avoid the excessive power of the leader to replace the message \(m\) with \(m^\prime\) and produce another challenge \(c^\prime\). Through this signature algorithm, the leader is hard to forge a joint signature \((c^\prime,S^\prime)\) by \(k\)-sum problem attacks.

\noindent\textbf{Key Verification.} Similar to Gamma signature, given an input as a public key \(pk\) as well as its corresponding proof of possession such that \(pk=(y,\pi)\), \(\pi=(a,d)\), the key verification algorithm \(\textsf{KVf}(par, pk)\) checks whether it satisfies that \(a=H_1(g_1,V)\), where \(V= (g_1^d y^b)^{a^{-1}}\)  and \(b= H_2(y)\). If not, the public key \(pk\) is invalid and must be discarded.

\noindent\textbf{Verification.} Given an input as a joint signature \(\sigma= (c,S)\) on an announcement \(m\) as well as the aggregated public key \(\tilde{X}\), \(\textsf{Vf}(par,\tilde{X},m,\sigma)\) computes \(e=H_3(m)\) and \(\tilde{V}=(g_1^S \tilde{X}^{e})^{c^{-1}}\), and then checks whether the equation satisfies \(c=H_0(g_1,\tilde{V},\tilde{X})\). If not, \((c,S)\) is an invalid signature. Otherwise, it is valid and the verifier accepts it.

\subsection{Advanced Gamma Multi-Signature Scheme}
From the signature construction of the proposed GMS, it can be seen that the generation of a collective challenge \(c\) has nothing to do with the announcement \(m\). Therefore, if the challenge \(c\) can be precomputed offline, we can change the running order of the above four phases in the signing algorithm to achieve better online performance. Meanwhile, we choose to run key aggregation algorithm in Commitment and Challenge phases, so that it can be executed distributedly. Therefore, the signing algorithm can be modified and optimized from GMS. We call this new scheme as Advanced Gamma Multi-Signature (AGMS).

In AGMS, we define Commitment and Challenge phases as pre-signing phases or offline signing, where each signer in a spanning tree structure comes to an agreement (challenge \(c\)) before the announcement \(m\) arrives. And then, Announcement and Response phases are defined as the formal-signing phases or online signing, where the leader receives the announcement \(m\) to be signed and produces the joint signature \(\sigma= (c,S)\). The signing algorithm in AGMS is described as follows.\\

\begin{figure}
\includegraphics[width=0.5\textwidth]{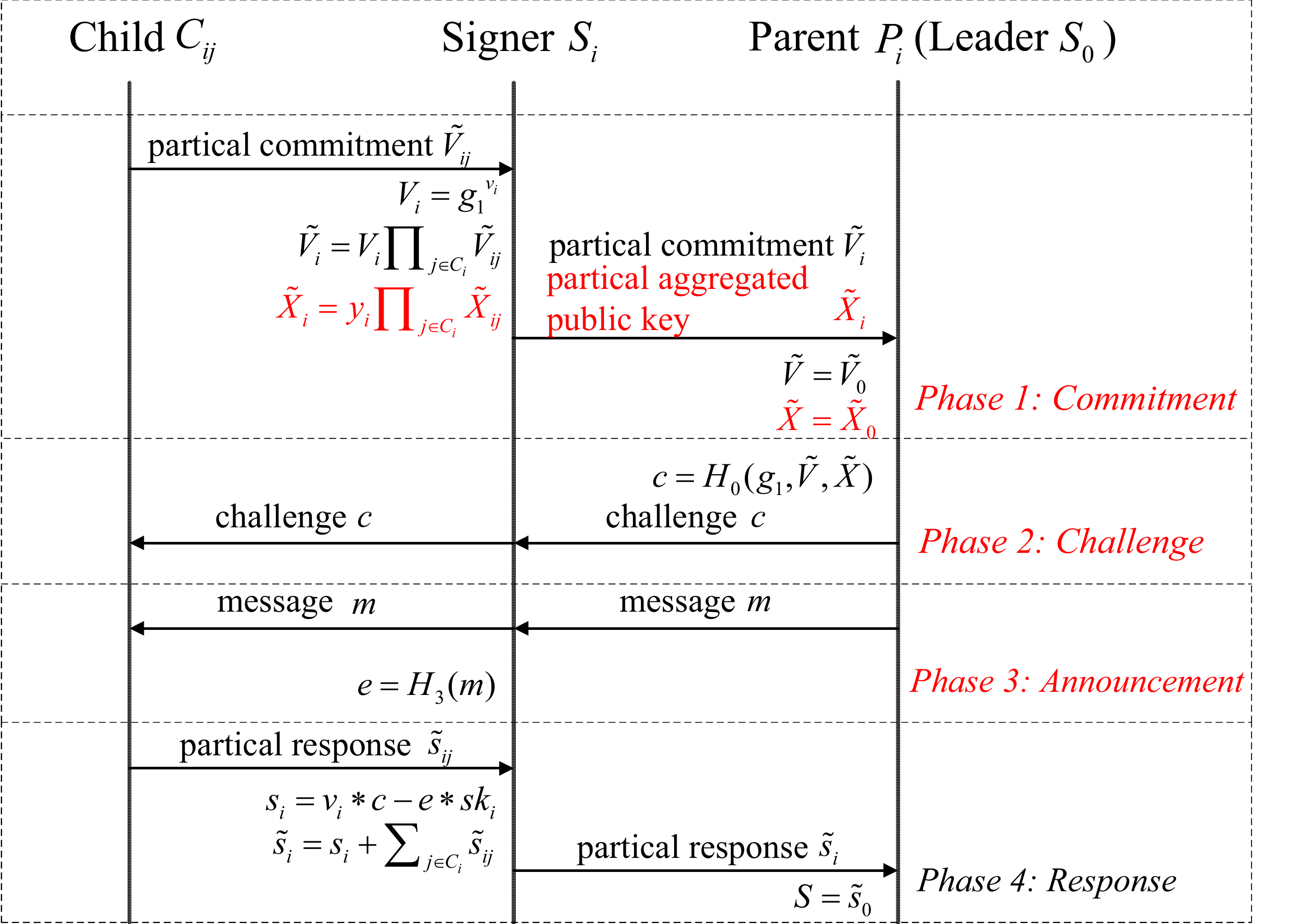}
\caption{The signing algorithm of the proposed AGMS scheme (Text in red indicates changes from Fig. 1. We suppose that signer \(S_i\) holds the key pair \((pk_i,sk_i)\), where \(pk_i=(y_i,\pi_i)\), and parent \(P_i\) works as a leader \(S_0\). If parent \(P_i\) is not a leader, it just works as signer \(S_i\). The key aggregation algorithm also runs together with the signing algorithm. Finally, the leader \(S_0\) outputs \((c,S)\) as the joint signature.)}
\label{AGMS}
\end{figure}

\noindent\textbf{Signing.} We also set \(C_i=\{C_{ij}\}\) as the set of children of one signer \(S_i\) in the spanning tree structure, and \(P_i\) as the parent of signer \(S_i\). Assume \(S_0\) to be the root of the tree, so called the leader. The signer \(S_i\) runs signing algorithm \(\textsf{Sign}(par,(pk_i, sk_i),m,\tau)\) in a tree \(\tau\) for four phases, which is shown in Fig. \ref{AGMS}.

\noindent\emph{\underline{Phase 1: Commitment.}} This process is run in a bottom-up way by each node \(S_i\). Specifically, given a node \(S_i\), choose a random secret value \(v_i\) and compute \(V_i=g_1^{v_i}\). Then, \(S_i\) waits for each immediate child \(j\)'s partial commitment \(\tilde{V}_{ij}\) and the partial aggregated public key \(\tilde{X}_{ij}\). When all the partial commitments are received, \(S_i\) computes \(\tilde{V}_i=V_i\prod_{j\in C_i} \tilde{V}_{ij}\) and \(\tilde{X}_i=y_i\prod_{j\in C_i}{\tilde{X}_{ij}}\). After that, the result \(\tilde{V}_i\) and \(\tilde{X}_i\) is sent to its parent \(P_i\) unless \(S_i\) is the leader (i.e. \(i=0\)).

\noindent\emph{\underline{Phase 2: Challenge.}} The leader \(S_0\) waits for each immediate child's partial commitment value \(\tilde{V}_{0j}\), partial aggregated public key \(\tilde{X}_{0j}\), and computes the final commitment \(\tilde{V}=\tilde{V}_0={V}_0\prod_{j\in C_0}\tilde{V}_{0j}\), as well as the aggregated public key \(\tilde{X}= \tilde{X}_0=y_0\prod_{j\in C_0}{\tilde{X}_{0j}}\). So, the collective challenge is \(c= H_0(g_1,\tilde{V},\tilde{X})\). The value \(c\), as a part of the joint signature, can be sent to the verifier in advance or stored at the leader. After that, the leader sends the shared challenge value \(c\) back to its children. All the signers \(S_i\) store \(c\) and precompute their own partial value \(v_i *c\).

\noindent\emph{\underline{Phase 3: Announcement.}} When the leader \(S_0\) receives a message \(m\), it starts to multicast the announcement \(m\) to its children top-down in the tree structure.

\noindent\emph{\underline{Phase 4: Response.}} When \(S_i\) receives announcement \(m\), it only computes \(e*sk_i\) and adds the previous partial value \(v_i *c\) to attain the individual response: \(s_i=v_i*c-e*sk_i\), where \(e=H_3(m)\). Then, it waits for each partial response \(\tilde{s}_{ij}\) from its immediate child \(j\). When all the partial responses are received, it sets \(\tilde{s}_i= s_i+\sum_{j\in C_i}{\tilde{s}_{ij}}\). After that, the result \(\tilde{s}_i\) is sent to its parent \(P_i\) unless \(S_i\) is the leader (i.e. \(i=0\)). Finally, the leader \(S_0\) computes the final response \(S= \tilde{s}_0=s_0+\sum_{j\in C_0}{\tilde{s}_{0j}}\) and outputs the joint signature \((c,S)\).

In summary, we have proposed two multi-signature schemes GMS and AGMS in this section. GMS focuses on the security improvement, where the verification algorithm for public key is deployed to defeat rogue-key attacks, and the signing algorithm is improved to resist $k$-sum problem attacks and avoid the leader modifying the message to produce another challenge. Meanwhile, the signing algorithm is split into online and offline parts. Furthermore, AGMS focuses on the efficiency improvement, where the running order of phases in signing algorithm is adjusted, and the key aggregation algorithm is executed distributedly, so as to obtain better online performance.

\section{Security Analysis}
In this section, we analyze security of the proposed AGMS scheme in details. In particular, security of a multi-signature scheme should satisfy two basic requirements.

First, a multi-signature scheme should be complete. That is, if we build up a system by \(\textsf{Pg}(\kappa)\), generate a set of public and private key pairs \((pk,sk)\) by \(\textsf{Kg}(par)\), and produce a joint signature \(\sigma\) on an announcement \(m\) representing a set of signers in a tree \(\tau\) by \(\textsf{Sign}(par,\mathcal{SK},m,\tau)\), then we should be able to use \(\tilde{X}\), generated from \(\textsf{KAg}(\mathcal{PK})\), to successfully output \(\textsf{KVf}(par,pk)=1\) and \(\textsf{Vf}(par,\tilde{X},m,\sigma)=1\). As these two verification equations are true, the proposed scheme AGMS satisfy the completeness requirement.

Second, a multi-signature scheme should be unforgeable. We prove that the proposed scheme AGMS can achieve unforgeability under current interactive attacks. The analysis is described as follows.

\noindent\emph{Lemma 1 (General forking lemma \cite{04DBLP:conf/ccs/BellareN06}):} Let \(\mathcal{C}\) be a randomized probabilistic algorithm. When given input \((x,h_1,\cdots,h_{q},\rho)\) with access to oracle \(\mathcal{O}\) of size \(\lambda\), where \(x\) is generated by the input generator \(\textsf{IG}\); \(\rho\) refers to \(\mathcal{C}\)'s random tape; \(h_1,\cdots,h_{q}\) are some random chosen values from \(\mathbb{Z}_q\); then \(\mathcal{C}\) outputs a pair \((J,y)\). Let \(\pi\) be the space of all the vectors \((x,h_1,\cdots,h_{q},\rho)\). Let \(acc\) be the probability that \(\mathcal{C}\) can successfully output \((J,y)\) when given inputs \((x,h_1,\cdots,h_{q},\rho)\), where \(J\) is a non-empty subsets of \(\{1,\cdots,q\}\).

For a given \(x\), the forking lemma algorithm \( \ F_{\mathcal{C}} (x)\) is described as follows.

\noindent{\(\ F_{\mathcal{C}} (x)\)}:

Pick a random tape \(\rho\) for \(\mathcal{C}\)

\(h_1,\cdots,h_{q}\leftarrow \mathcal{O}\)

\((J,y)\leftarrow\mathcal{C}(x,h_1,\cdots,h_{q},\rho)\)

if \(J=0\) then

\qquad return \((0,\perp,\perp)\)

\(h_1^\prime,\cdots,h_q^\prime\leftarrow \mathcal{O}\)

\((J^\prime,y^\prime)\leftarrow\mathcal{C}(x,h_1^\prime,\cdots,h_q^\prime,\rho)\)

if \(J=J^\prime\) and \(h_J\neq h^\prime_{J^{\prime}}\)

\qquad return \((1,y,y^\prime)\)

else

\qquad return \((0,\perp,\perp)\)

We let \(frk\) be the probability that \(F_\mathcal{C}\) successfully outputs \((1,y,y^\prime)\) as shown below:
\begin{equation}
frk=\textup{Pr}[b=1:x\leftarrow \textsf{IG};(b,y,y^\prime)\leftarrow F_
\mathcal{C}
(x)] \ .
\end{equation}

 So that we have:
\begin{equation}
frk\geq acc(\frac{acc}{q} - \frac{1}{\lambda}) \ .
\end{equation}

\noindent\emph{Lemma 2:} Let \(\prod =(\textsf{Pg},\textsf{Kg},\textsf{KAg},\textsf{Sign},\textsf{KVf},\textsf{Vf})\) be a multi-signature scheme. We define the security of a multi-signature scheme as the universal unforgeability under a chosen message attack against a set of honest players. We can say CoSi is \((t,q_s,q_f,N,\varepsilon)\)-secure in the random-oracle model, if given \(N\) as the maximum number of participating signers that the adversary needs to run at most \(t\) time, with the probability of forgeability of at least \(\varepsilon\), making at most \(q_s\) signature queries as well as \(q_f\) random oracle queries.

As CoSi is based on Schnorr signature, we follow the random oracle model. In CoSi, we only assume \(H_0(\tilde{V},m)\) are modeled as random oracles, which is only \(t_f,\varepsilon_{cr}\)-collision resistant, so that we may prove CoSi secure in the random oracle under the discrete algorithm assumption. Differently, as for AGMS, we follow the so-called general key registered model \cite{24DBLP:conf/scn/BagherzandiJ08}, where the validity of each public key must be checked by the signature verifier.
In the proposed scheme AGMS, we only assume
\(H_0(g_1,\tilde{V},\tilde{X}):{\{0,1\}}^{\ast}\rightarrow {\{0,1\}}^{\kappa}\) and \(H_1(g_1,u^\ast):{\{0,1\}}^{\ast}\rightarrow {\{0,1\}}^{\kappa}\) are modeled as random oracles, and define the other two hash functions \(H_2(y^\ast):{\{0,1\}}^{\ast}\rightarrow {\{0,1\} }^{\kappa}\) and \(H_3(m):{\{0,1\}}^{\ast}\rightarrow {\{0,1\}}^{\kappa}\) as target one-way hash functions, which are \((t_f,\varepsilon_{cr})\)-collision resistant and \((t_f,\varepsilon_{tow})\)-target one-way, to mitigate the dependency of provable security on random oracles.

\noindent\emph{Theorem 1:} Suppose that AGMS is \((t^\prime,q_s,q_f,N,\varepsilon^\prime)\)-secure under the discrete logarithm problem, there exists an algorithm \(\mathcal{C}\) that if we take uniformly random group elements \(X^\ast\), two uniformly random chosen \(\kappa\)-bit strings \(H_0\), \(H_1\) for a total of \((q_s+q_f)\) times and two target one-way \(\kappa\)-bit strings \(H_2\), \(H_3\) as inputs, then, \(\mathcal{C}\) can successfully output a tuple \((i_0,i_3,\mathcal{PK},S,i_1,i_2)\), satisfying \(\tilde{X}=\prod_{pk_i\in \mathcal{PK}} pk_i\) and  \(H_0(g_1,(g_1^{S}\tilde{X}^{i_3})^{i_0^{-1}},\tilde{X})=i_0\). Here, \(i_0\in (H_{01},\cdots, H_{0 q})\), \(i_1\in (H_{11},\cdots,H_{1 q})\), and \(i_2\), \(i_3\) are the two target one-way hash values involved in the corresponding set of signers' public keys \(\mathcal{PK}\). Assume that \(N\) is the maximum number of signers that participate in AGMS. Then, the running time of algorithm \(\mathcal{C}\) is at most \(t^\prime\), and algorithm \(\mathcal{C}\) succeeds with the probability of \(\varepsilon^\prime\) such that
\begin{equation}
\varepsilon^\prime\geq acc(\frac{acc}{q_s+q_f}-\frac{1}{2^{\kappa}})-\varepsilon_{tow} \ ,\end{equation}
where
\begin{equation}
acc\geq(1-\frac{q_s(2q_f+q_s-1)}{2^{3\kappa+1}})(\varepsilon-\frac{N+1}{2^{\kappa}}-\frac{N(N-1)+2}{2}\varepsilon_{cr}) \ .
\end{equation}

\noindent\emph{Proof:} We construct a four-stage game for an algorithm $\mathcal{C}$ around a $(t,q_s,q_f,N,\varepsilon)$-forger $\mathcal{F}$. Assume that the involved signers behave honestly. Given the random public key set \(\mathcal{PK}=(pk_1,\cdots,pk_N)\), we simulate the game in the following steps.

\noindent\textbf{Setup:} Algorithm \(\mathcal{C}\) initializes \(par=(\mathbb{G},g_1,q)\leftarrow \textsf{Pg}(\kappa)\), \((pk,sk)\leftarrow \textsf{Kg}(par)\), and two empty hash query sets \(S_{H_0}\) and \(S_{H_1}\), corresponding to the queries of \(H_0\) and \(H_1\) respectively such that \((d_{T1},\cdots, d_{Tq_s},d_{T(q_s+1)},\cdots,d_{T(q_s+q_f)})\leftarrow (\{0,1\}^{\kappa})^{q_s+q_f}\), \((T=0,1)\). Then, we construct a ``proof of possession" of \(sk_i\). \(\mathcal{C}\) provides a random tape \(\rho\) to \(\mathcal{F}\), and runs $\mathcal{F}$ as a signer with the public key $pk_1=(y_1,\pi_1)$.

\noindent\textbf{RO queries:} As for CoSi, there only involves one hash value that consists of the final commitment \(\tilde{V}\) and a message \(m\). Differently, in the proposed AGMS, there are two independent hash values to query. For each query set, under the \(i\)-th query \((1\leq i\leq q_f)\) denoted by \(Q_{Ti},(T=0,1)\) from \(\mathcal{F}\), \(\mathcal{C}\) firstly checks whether the value \(Q_{Ti}\) has been defined before. If yes, \(\mathcal{C}\) gives up the repeated value \(H_T(Q_{Ti})=\alpha\). Otherwise, \(\mathcal{C}\) defines \(H_T(Q_{Ti})=d_{T(q_s+i)}\), stores the record \((j=q_s+i,Q_{Ti},H_T(Q_{Ti})=d_{T(q_s+i)})\) in the corresponding set \(S_{H_T}(T=0,1)\) and then sends the values \(d_{T(q_s+i)}\) to \(\mathcal{F}\).

\noindent\textbf{Signature queries:} With the set of public keys \(\mathcal{PK}=(pk_1,\cdots,pk_N)\) and some messages \(m\), \(\mathcal{C}\) firstly simulates each self-signed information \(\pi^\ast=(d^\ast,w^\ast)\) by randomly selecting two values \(d^\ast,w^\ast\stackrel{\$}\leftarrow\mathbb{Z}_q\), and then computing \(u^\ast= (g_1^{w^\ast}y^{\ast b^\ast})^{d^{\ast -1}}\), where \(b^\ast= H_2(y^\ast)\). On input \(pk^\ast=(y^\ast,\pi^\ast)\) with the random tape \(\rho\), \(\mathcal{C}\) makes the query \( H_1(g_1,u^\ast)=d^\ast\). When there exists \(H_1(g_1,u_i^\ast)\) that is never defined in previous queries, \(\mathcal{C}\) sets \(H_1(Q_{1i})=d_{1i}\) and stores \((j=i,Q_{1i},H_1(Q_{1i})=d_{1i})\) in the set \(S_{H_1}\). After receiving \(\tilde{X}=\prod_{pk_i\in \mathcal{PK}} pk_i\) and \(\tilde{V}=\prod_{i=1}^N V_i\) from \(\mathcal{C}\), \(\mathcal{F}\) simulates a query \(c= H_0(g_1,\tilde{V},\tilde{X})\) that is never defined before, stores \((j=i,Q_{0i},H_0(Q_{0i})=d_{0i})\) in the set \(S_{H_0}\) and sends \(c\) to its children without knowing the message \(m\).
\(\mathcal{C}\) can return partial queries value \(c=H_0(Q_{0i})=d_{0i}\) firstly. This is hard for some schemes including CoSi to produce the partial signature value in advance.
Finally, after knowing the message \(m\), similar to signer \(S_i\), \(\mathcal{C}\) waits for the response \(\tilde{s}_{ij}\) that comes from its children \(j\in C_i\), proceeds to compute and send \(\tilde{s_i}= s_i+\sum_{j\in {C}_i} \tilde{s}_{ij}\ mod\ q\) to its parent, where \(s_i=v_i *c-e*sk_i\ mod\ q\). Finally, \(\mathcal{C}\) returns \((c,S)\) as the joint signature.

We assume that there are several cases that may happen and cause \(\mathcal{C}\) to abort the execution. (1) The value \(Q_{0j}\leftarrow(g_1,\tilde{V}_j,\tilde{X}_j)\) that \(\mathcal{F}\) can successfully guess is equal to \(Q_{0i}\leftarrow(g_1,\tilde{V}_i,\tilde{X}_i)\) that is already defined before. (2) \(\mathcal{F}\) successfully attains the value \(Q_{0j}\leftarrow(g_1,\tilde{V}_j,\tilde{X}_j)\) that is never defined before by the birthday paradox. If either of the two cases happens, we set \(bad\leftarrow true\).

\noindent\textbf{Output:} Eventually, $\mathcal{F}$ outputs a forged multi-signature $(c^\prime,S^\prime)$ on the message $m^\prime$ for a multiset $\mathcal{PK}^\prime$. Without loss of generality, we assume the following conditions. (1) All hash queries that are involved in the verification of the forgery; and the proof of possession in \(\mathcal{PK}^\prime\) are made and recorded in sets \(S_T(T=0,1)\). (2) There do not exist any two different values \(Q_{2i}\) and \(Q_{2j}\) in \(\mathcal{PK}^\prime\) such that \(H_2(Q_{2i})=H_2(Q_{2j})=\alpha\). (3) There do not exist any two different values \(Q_{3i}\) and \(Q_{3j}\) such that \(H_3(Q_{3i})=H_3(Q_{3j})=\alpha\). When \(\mathcal{F}\)'s forgery is verified to be true, algorithm \(\mathcal{C}\) halts and returns 
\((J,(c^\prime,e^\prime,S^\prime,\mathcal{PK}^\prime))\).
If not, algorithm \(\mathcal{C}\) returns \((0,\perp)\) and fails to forge a joint signature.

As we defined above, \(H_T\stackrel{\$}\leftarrow {\{0,1\}}^{\kappa}\) \((T=2,3)\) is a \((t_f,\varepsilon_{tow})\)-target one-way as well as \((t_f,\varepsilon_{cr})\)-collision resistant hash function. Let \(t_s\) denote the running time of a signing query and \(t_{ex}\) denote the running time of extracting \(\mathcal{SK}\) using the generalized forking lemma \(F_\mathcal{C}\). Based on the above description, we can derive that: the event \(bad\leftarrow true\) happens with the probability of
\(\textup{Pr}(bad\leftarrow true)\leq\frac{q_sq_f}{2^{3\kappa}}+\frac{q_s(q_s-1)}{2^{3\kappa+1}}\).
Considering that the event \(bad\leftarrow true\) does not happen, the probability that \(\mathcal{C}\) successfully outputs a forged signature \((c^\prime\,S^\prime)\) satisfying the above requirements is \(acc\geq(1-\frac{q_s(2q_f+q_s-1)}{2^{3\kappa+1}})(\varepsilon-\frac{N+1}{2^{\kappa}}-\frac{N(N-1)+2}{2}\varepsilon_{cr})\). Then, as \(\mathcal{F}\) is described above, algorithm \(\mathcal{C}\) is \((t^\prime,\varepsilon^\prime)\)-break the hash property of one-wayness, where the running time is at most 
\(t^\prime=(2N+2)t_f+(2N+2)q_st_s+t_{ex}+O((N+1)q_f)\), and equation (3) and equation (4) are true.

We further prove the theorem in more details by constructing an algorithm \(\mathcal{C}^\prime\). Suppose there is an algorithm $\mathcal{C}^\prime$, given input group elements $X^\ast$ and a signature forger $\mathcal{F}$ that is the same as described above, $\mathcal{C}^\prime$ can solve the discrete logarithm problem in $\mathbb{G}$. Finally, \(\mathcal{C}^\prime\) successfully outputs a forgery, using \(F_\mathcal{C}\) defined in \noindent\emph{Lemma 1}. \(\mathcal{C}^\prime\) proceeds as follows.

We set \((1,(c,e,S,N)\) and \((1,(c^\prime,e^\prime,S^\prime,N^\prime)\) as two different outputs of \(\mathcal{C}^\prime\) associated with the forgery such that:
\begin{center}
\(g_1^S=\tilde{V}^c\tilde{X}^{-e}=\tilde{V}^c\prod_{i=1}^N y_i^{-e}\) and \(g_1^{S^\prime}=\tilde{V}^{\prime c^\prime}\tilde{X}^{\prime -e^\prime}=\tilde{V}^{\prime c^\prime}\prod_{i=1}^{N^\prime} y_i^{\prime -e^\prime}\) ,
\end{center}
where we set \(\mathcal{PK}={(pk_1,\cdots,pk_N)}\) and \(\mathcal{PK}^\prime={(pk_1^\prime,\cdots,pk_{N^\prime}^\prime)}\) as two sets of public keys involved in \(\mathcal{F}\)'s forgery. According to the construction of \(\mathcal{C}^\prime\), we should hold \(\tilde{V}^\prime=\tilde{V}\), \(c^{\prime -1}e^{\prime}\neq c^{-1}e\), \(N^\prime=N\) and \(y_i^\prime=y_i(1\leq i \leq N)\). Therefore, we have that:
\begin{equation}
\tilde{V}=g_1^{Sc^{-1}}\prod_{i=1}^{N} y_i^{c^{-1}e},
\end{equation}
and
\begin{equation}
\tilde{V}=g_1^{S^\prime c^{\prime-1}}\prod_{i=1}^{N} y_i^{c^{\prime-1} e^\prime}.
\end{equation}
Based on equations (5) and (6), it will yield:
\begin{equation}
g_1^{Sc^{-1}-S^\prime c^{\prime -1}}=\prod_{i=1}^N y_i^{c^{\prime -1}e^\prime -c^{-1}e}=\prod_{pk_i\in \mathcal{PK}}(y_i)^{c^{\prime -1}e^\prime -c^{-1}e }.
\end{equation}
Because
 \begin{equation}
 \tilde{X}=\prod_{y_i\in \mathcal{PK}} pk_i=g_1^{\sum_{pk_i\in \mathcal{PK}}sk_i},
 \end{equation}
\(\mathcal{C}^\prime\) can successfully attain the discrete logarithm of \(pk_1\) as
\begin{center}
\(\frac{Sc^{-1}-S^\prime c^{\prime -1}}{c^{\prime -1}e^\prime  -c^{-1}e}-\sum_{pk_i\in \mathcal{PK} \setminus pk_1}sk_i\ mod\ q\) .
\end{center}
That is, if $c^\prime=c$ and $e^\prime\neq e$,the forger $\mathcal{F}$ can successfully extract all $\mathcal{SK}$ except its own $sk_1$. Using \noindent\emph{Lemma 1}, we can compute that the probability for forger \(\mathcal{F}\) to obtain two different outputs, where \(c^\prime\neq c\) or \(e^\prime\neq e\), is \(frk \geq acc(\frac{acc}{q_f+q_s}-\frac{1}{2^{\kappa}})\). Thus,
the probability of \(\mathcal{C}^\prime\) in doing so is given as \( frk \geq acc(\frac{acc}{q_f+q_s}-\frac{1}{2^{\kappa}})-\varepsilon_{tow}\), where \(acc\) satisfies equation (4). The total running time of algorithm \(\mathcal{C}^\prime\) is at most that of \(F_\mathcal{C}\) plus \(O(N)\) operations.
In other words, the proposed scheme AGMS can achieve unforgeability under current interactive attacks.

\section{Performance Analysis}
\subsection{Theoretical Analysis}
In theory, we briefly compare the proposed schemes GMS and AGMS with current most popular multi-signature schemes, including BN \cite{04DBLP:conf/ccs/BellareN06}, CoSi \cite{02DBLP:conf/sp/SytaTVWJGGKF16}, and Musig \cite{DBLP:journals/dcc/MaxwellPSW19}.

The property comparisons of these schemes are summarized in Table \ref{fig11}.
First, based on the prototype of Schnorr signature, these schemes are proved to be standard existential unforgeable under the adaptive chosen message attacks. However, BN, CoSi, and Musig involve only one hash value that consists of the random value and message, meaning that these schemes do not support to precompute the partial signature \(c\) and are possible to be forged by \(k\)-sum problem attacks. Therefore, it is uncertain whether they can be proved secure in concurrent interactive protocols.
Differently, the proposed GMS and AGMS are based on Gamma signature, which involve two different independent hash values, and thus can be secure against \(k\)-sum problem attacks. As the proposed GMS produces the challenge \(c\) after the message \(m\) comes, only the proposed AGMS can achieve provable security in concurrent interactive protocols. That is to say, the leader in AGMS can work as a representative of a group of signers in a spanning tree structure, precompute the challenge \(c\), and achieve two-round interactive telecommunications with other individuals or groups in a secure way. AGMS is also the only scheme that can support the partial signature value \(c\) to be public.
Second, as mentioned before, CoSi is easily to be forged by rogue-key attacks. BN and Musig added one more round protocol to exchange their individual commitments to other signers, which is a solution to avoid rogue-key attacks. But this approach inevitably leads to high communication and computation overhead. The two proposed schemes, GMS and AGMS, use proof of possession, which is an efficient way to avoid rogue-key attacks.
Third, with the spanning tree structure, GMS, AGMS, and CoSi can reach high scalability, which is hard-to-reach by BN or Musig.
\begin{table*}[htb]
\caption{Properties of several multi-signature schemes}
\centering
\begin{tabularx}{\linewidth}{p{5cm}XXXXX}
\toprule Multi-signature schemes&Proposed GMS&Proposed AGMS&BN&CoSi&Musig\\
\midrule
Provable security (Standard)&Yes&Yes&Yes&Yes&Yes\\
Provable security (Concurrent interactive)&Uncertain&Yes&Uncertain&Uncertain&Uncertain\\
Support challenge \(c\) public&No&Yes&No&No&No\\
Against rogue-key attacks&Yes&Yes&Yes&No&Yes\\
Against \(k\)-sum problem attacks&Yes&Yes&No&No&No\\
Rounds&2&2&3&2&3\\
Spanning tree structure&Yes&Yes&No&Yes&No\\
\bottomrule
\end{tabularx}
\label{fig11}
\end{table*}

\begin{table*}[htb]
\caption{Efficiencies of several multi-signature schemes}
\centering
\begin{tabularx}{\linewidth}{p{5cm}XXXXX}
\toprule Multi-signature schemes&Proposed GMS&Proposed AGMS&BN&CoSi&Musig\\
\midrule
\(\textsf{KAg}\)&- &- &- &- &1\(\cdot\)exp$^N$\\
\(\textsf{Sign}\) (online signing)&1\(\cdot\)exp&- &1\(\cdot\)exp&1\(\cdot\)exp&1\(\cdot\)exp\\
\(\textsf{Sign}\) (offline signing)&-&1\(\cdot\)exp&- &- &- \\
\(\textsf{Vf}\)&1\(\cdot\)exp$^3$&1\(\cdot\)exp$^3$&1\(\cdot\)exp$^{N+1}$&1\(\cdot\)exp$^2$&1\(\cdot\)exp$^2$\\
\(\textsf{KVf}\)&1\(\cdot\)exp$^3$&1\(\cdot\)exp$^3$&- &- &- \\
Total \((\textsf{Sign}+\textsf{Vf})\)&1\(\cdot\)exp$^{N+3}$&1\(\cdot\)exp$^{N+3}$&1\(\cdot\)exp$^{2N+1}$&1\(\cdot\)exp$^{N+2}$&1\(\cdot\)exp$^{N+2}$\\
Signature domain&\(\mathbb{Z}_q^2\)&\(\mathbb{Z}_q^2\)&\(\mathbb{G}\times \mathbb{Z}_q\)&\(\mathbb{Z}_q^2\)&\(\mathbb{G}\times \mathbb{Z}_q\)\\
\(pk\) domain&\(\mathbb{G}\times \mathbb{Z}_q^2\)&\(\mathbb{G}\times \mathbb{Z}_q^2\)&\(\mathbb{G}\)&\(\mathbb{G}\)&\(\mathbb{G}\)\\
\(\tilde{X}\) domain&\(\mathbb{G}\)&\(\mathbb{G}\)&\(\mathbb{G}^N\)&\(\mathbb{G}\)&\(\mathbb{G}\)\\
Offline storage&\(\mathbb{G}\times \mathbb{Z}_q\)&\(\mathbb{G}^2\)&\(\mathbb{G}\times \mathbb{Z}_q\)&\(\mathbb{G}\times \mathbb{Z}_q\)&\(\mathbb{G}\times \mathbb{Z}_q\)\\
\bottomrule
\end{tabularx}
\begin{tablenotes}
\small
\item (``-" denotes no exponentiation. ``exp" denotes an exponentiation. ``exp$^k$" denotes an $k$-multi-exponentiation in a group ``\(\mathbb{G}\)". ``\(N\)" denotes the number of signers involved in a multi-signature scheme.)
\end{tablenotes}
\label{fig22}
\end{table*}

Furthermore, we compare the efficiencies of these multi-signature schemes in Table \ref{fig22}.
In particular, Musig needs a very time-consuming \(\textsf{KAg}\) algorithm to construct a more secure joint signature without revealing individual signer's public key.
In the \(\textsf{Sign}\) algorithm, due to the advantage that the challenge \(c\) can be precomputed offline, the proposed AGMS performs better in online signing than all other schemes.
In the \(\textsf{Vf}\) algorithm, the proposed schemes GMS and AGMS require one more exponentiation when compared to BN and Musig.
Because of proof of possession, the two proposed schemes GMS and AGMS also require \(\textsf{KVf}\) algorithm against rogue-key attacks.
The total computation of the \(\textsf{Sign}\) and \(\textsf{Vf}\) algorithms in these two schemes is only slightly higher than that of CoSi and Musig, but much less than that of BN.
In the signature domain and \(\tilde{X}\) domain, the two proposed schemes require the smallest space among these multi-signature schemes. Only the \(pk\) domain needs more space than other schemes due to the proof of possession. In the offline storage, we can suggest that the signer in other schemes except AGMS to precompute and store \((v_i,V_i)\). But the signer in AGMS can store \((v_i,c)\), meaning that in terms of offline storage, AGMS only needs \(\mathbb{G}^2\), which is much smaller than \(\mathbb{G}\times \mathbb{Z}_q\) required by other schemes.
In summary, the proposed GMS and AGMS schemes are comparable to others in terms of efficiency, but AGMS enjoys the greatest efficiency in online signing and the smallest space in offline storage, which can avoid the network congestion and is suitable to be applied in real-time communications.

\subsection{Experimental Analysis}
In this subsection, 32 physical machines that consist of an Intel (R) Core (TM) i7-4790 processor and a RAM with total memory of 8GB are adopted for testing purpose. We implement the following schemes through Go\footnote[1]{http://golang.org/, January, 2015.} programming language. We use hash function SHA-512 \cite{U2012Secure} and SHA-512 based target one-way hash function \cite{01DBLP:journals/tifs/YaoZ13}. We run each experiment for 20 times and show the average results. As the experiment results have significant differences, to show every value, $y$-axis in Fig. [3]-[8] and Fig. [10]-[13] has logarithmic scale.

According to the difficulty assumptions and basic signature algorithms, we test RSA based multi-signature \cite{DBLP:conf/eurocrypt/HohenbergerW18}, BLS based multi-signature \cite{07DBLP:conf/asiacrypt/BonehDN18}, and Schnorr based multi-signature, including CoSi \cite{02DBLP:conf/sp/SytaTVWJGGKF16} and AGMS.
As Gamma signature, the basis of AGMS, is modified from Schnorr signature, and still based on the discrete logarithm problem, AGMS is classified to Schnorr based multi-signature schemes.
These experiments import two Go programming libraries: crypto\footnote[2]{Go cryptography libraries.} and pbc\footnote[3]{https://github.com/Nik-U/pbc, accessed December, 2018.}.
For the same security level, we define the elliptic curve is NIST P-224, and modulus for RSA is 2048-bit.
Through experiments, we have validated that the signature lengths for RSA, BLS, and Schnorr based schemes are 2048 bits, 224 bits, and 448 bits, respectively, indicating that BLS based signatures will take up the smallest storage space. However, as shown in Fig. 3, BLS based signature scheme takes significantly longer running time than the other two categories for both the signing and verification processes, as the bilinear pairing operation is time-consuming. On the other hand, although the time cost for verification algorithm of RSA based multi-signature is low, the total time is very close to that of Schnorr-based schemes (e.g. CoSi, AGMS). In addition, its signature length (2048 bits) will significantly increase the system storage overhead, and is usually unacceptable. With a reasonable signature length (448 bits), the experiment results validate that CoSi and AGMS yield the shortest running time for the signing process and a reasonable running time for the verification process.
Hence, the Schnorr based multi-signature schemes CoSi and AGMS are beneficial for achieving the balance of computational complexity and required storage space.

\begin{figure*}[htbp]
\begin{minipage}[t]{0.45\linewidth}
\centering
\includegraphics[width=\textwidth]{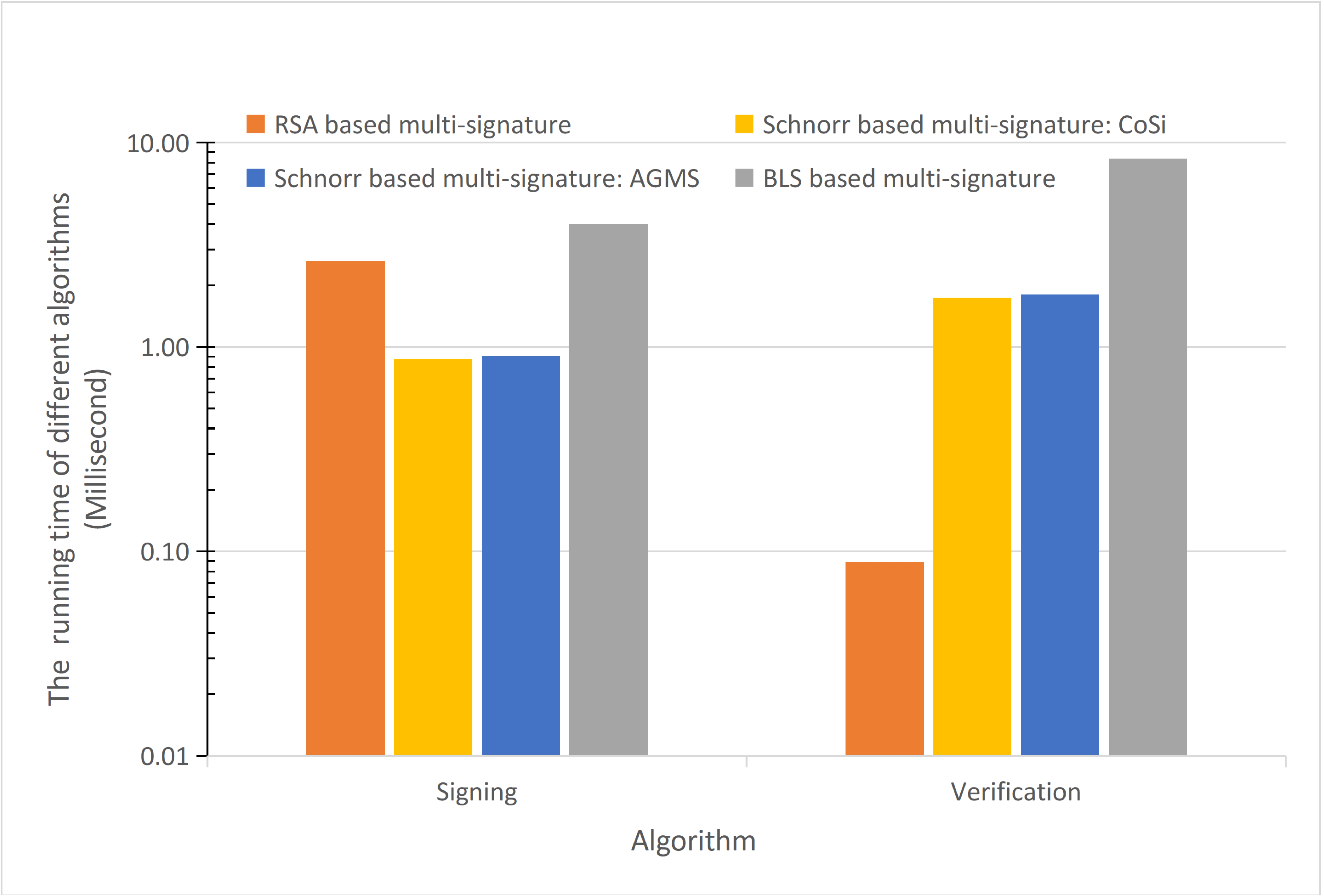}
\caption{The running time of signing and verification algorithms of typical difficulty assumptions based multi-signature schemes ($y$-axis has logarithmic scale.)}
\label{figassumption}
\end{minipage}%
\hfill
\begin{minipage}[t]{0.45\linewidth}
\centering
\includegraphics[width=\textwidth]{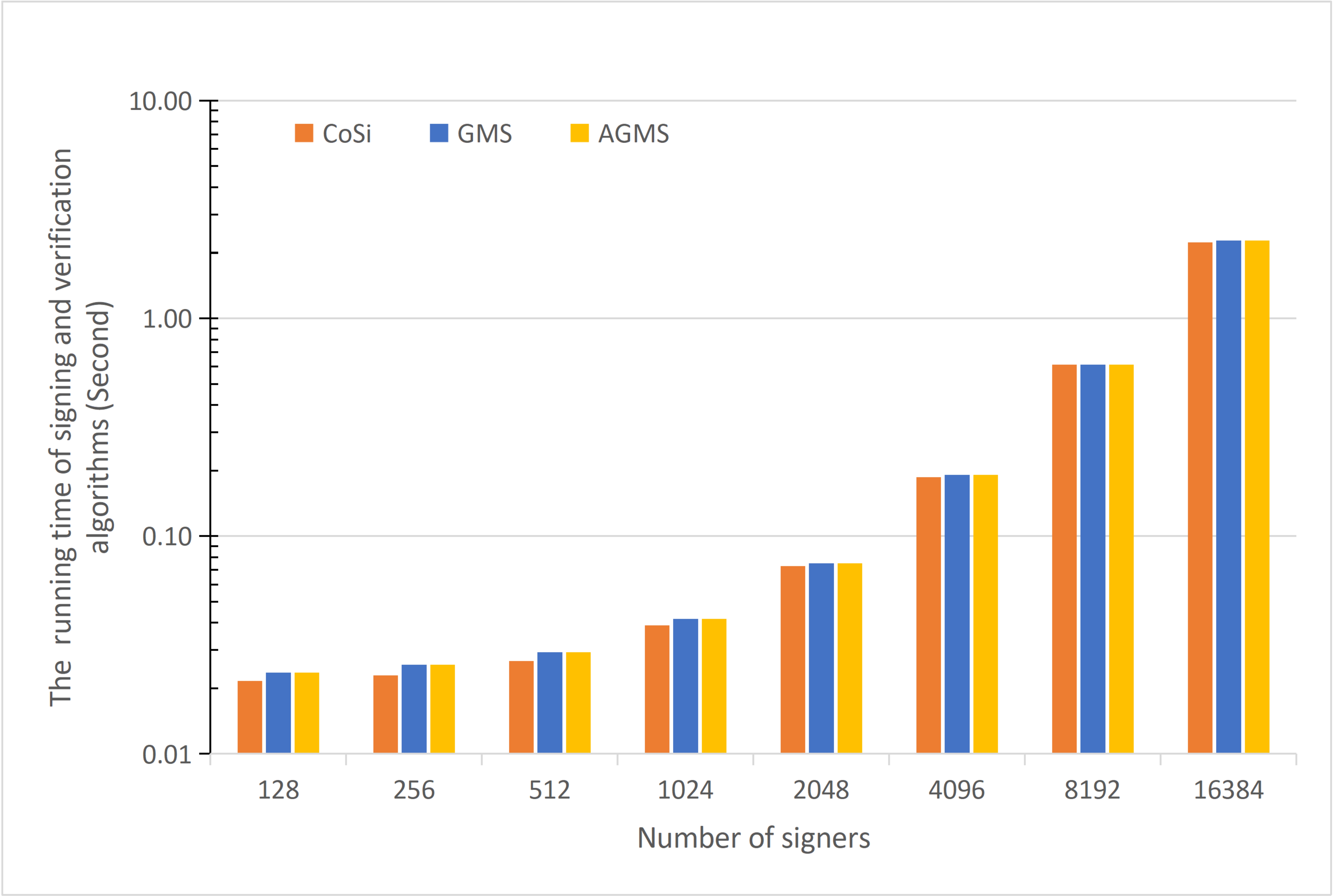}
\caption{The total CPU running time of signing and verification algorithms for CoSi, GMS, and AGMS. (The three algorithms achieve similar CPU running time, showing that the additional security features of the proposed algorithms do not sacrifice algorithm efficiency. $y$-axis has logarithmic scale.)} \label{fig3}
\end{minipage}%
\end{figure*}

Next, the two proposed schemes and CoSi are evaluated with a total amount of signers ranging from 128 to 16384, and all the signing nodes are created and connected in a tree structure. As the random depth of tree may influence the results, we set the tree depth to 3 and choose the branching factor according to the number of signers so as to keep it manageable. These experiments import two Go programming libraries: cothority\footnote[4]{https://github.com/dedis/cothority, accessed February, 2018.} and onet\footnote[5]{https://github.com/dedis/onet, accessed February, 2018.}. These schemes are based on elliptic curve 25519, and we ignore the computation time of key aggregation algorithm. From Fig. \ref{fig3}, we can find that, the offset among the total running time of signing and verification algorithms for these schemes is very close when the number of signers is up to 16384. The results confirm that the proposed schemes can easily scale up to thousands of signers as well.

\begin{figure*}[htbp]
\begin{minipage}[t]{0.45\linewidth}
\centering
\includegraphics[width=\textwidth]{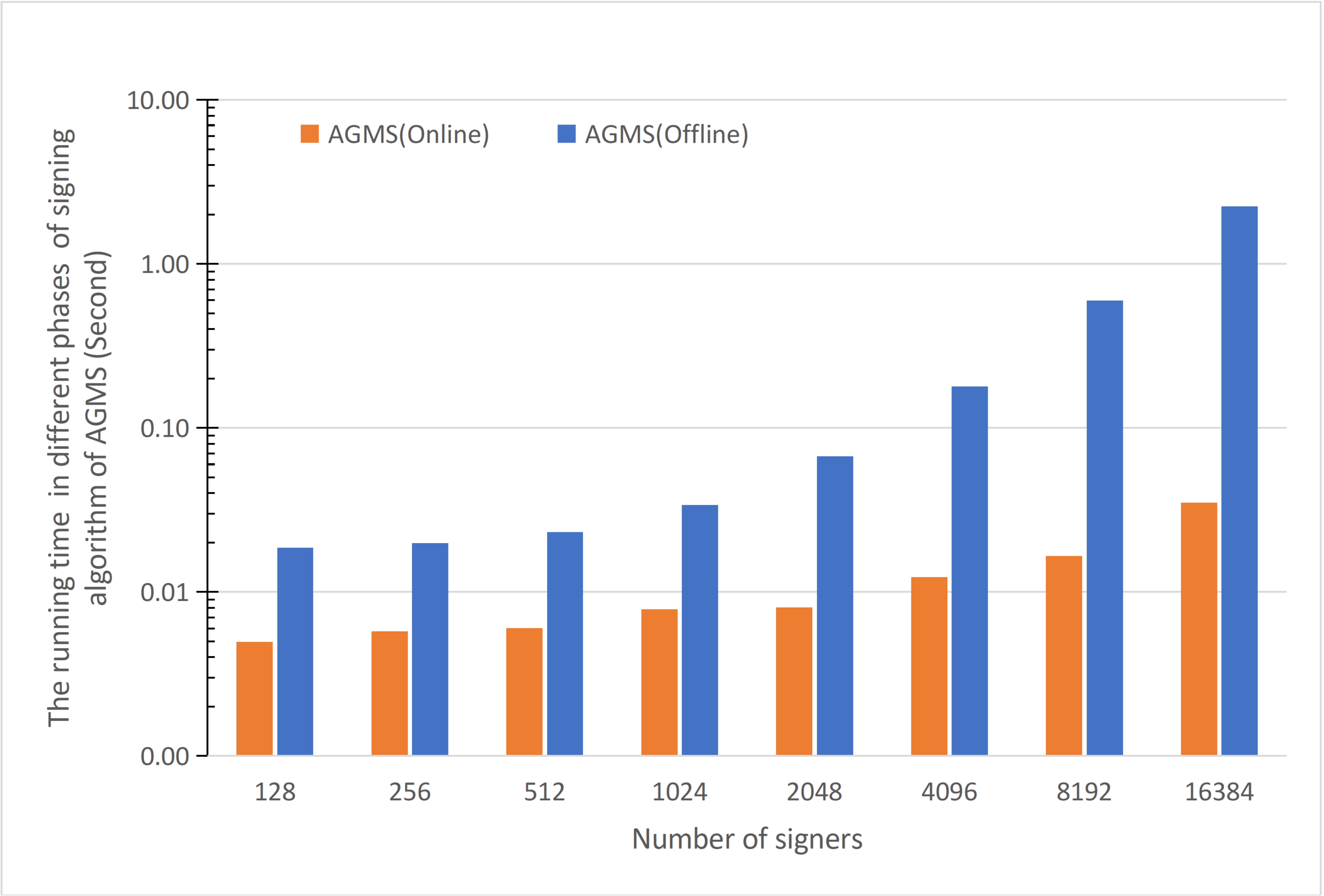}
\caption{The total CPU running time in different phases of signing algorithm in AGMS ($y$-axis has logarithmic scale.)} \label{fig4}
\end{minipage}
\hfill
\begin{minipage}[t]{0.45\linewidth}
\centering
\includegraphics[width=\textwidth]{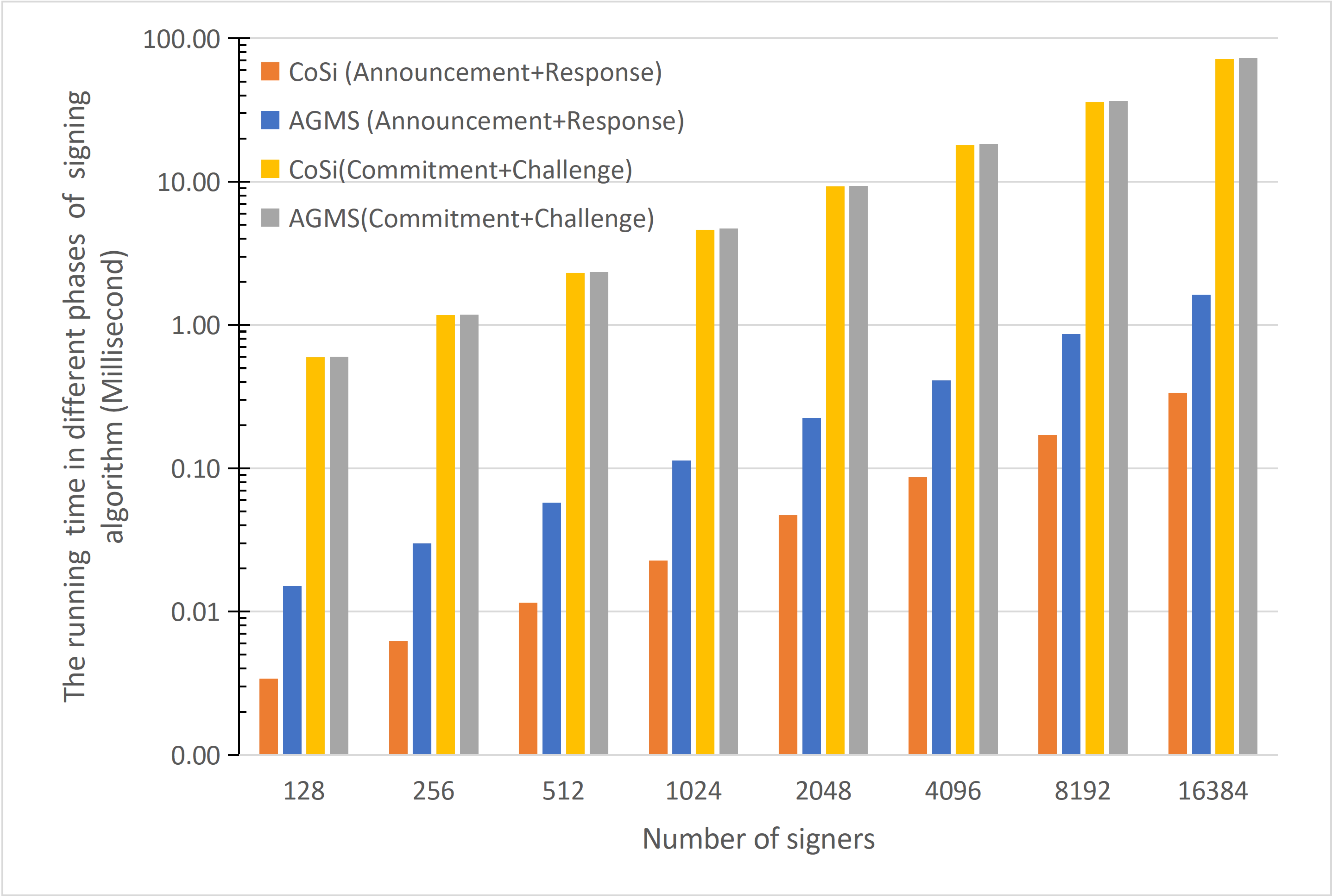}
\caption{The CPU running time on a leader node of CoSi and AGMS in different phases of signing algorithm ($y$-axis has logarithmic scale.)} \label{fig5}
\end{minipage}
\end{figure*}

Then, we test the running time of online signing phase and offline signing phase in the proposed AGMS. In the proposed AGMS, the online signing phase consists of the Announcement and Response phases, and the offline signing phase consists of the Commitment and Challenge phases. All the configurations remain the same as those in the first experiment. From the first experiment, we see that the total running time of signing algorithm of AGMS is very close to that of CoSi. As the offline signing phase needs a large amount of elliptic curve exponentiations, it accounts for the vast majority of the total running time of signing algorithm in AGMS. Therefore, the online signing part of the proposed AGMS scheme is very fast. When the number of signers goes up to 16384, we can find that the online signing time of AGMS is less than 1 second, accounting for only about 1\% of total running time of signing algorithm. Fig. \ref{fig4} depicts the results.

\begin{figure*}
\begin{minipage}[t]{0.45\linewidth}
\centering
\includegraphics[width=\textwidth]{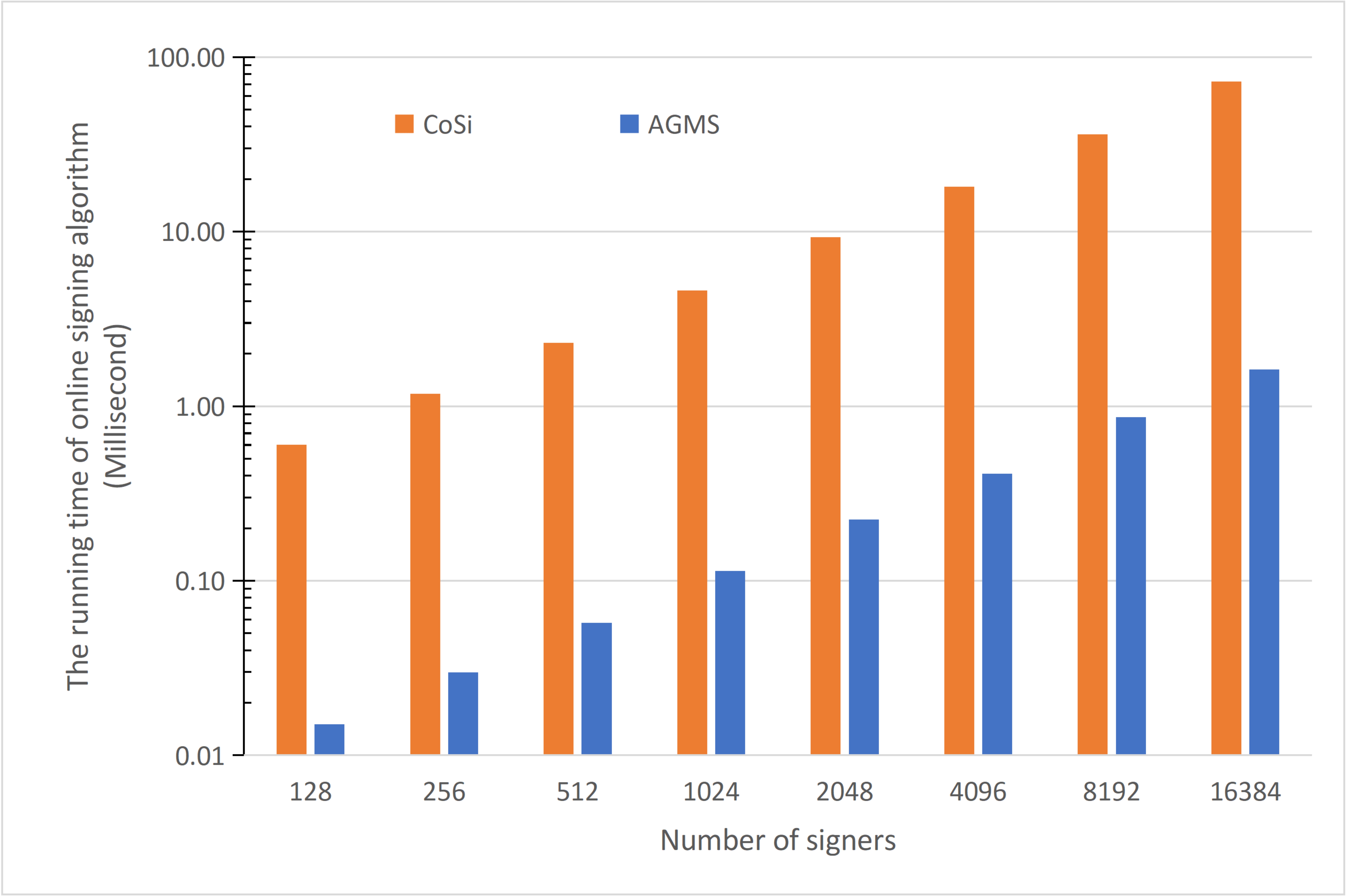}
\caption{The CPU running time on a leader node of CoSi and AGMS in online signing phase ($y$-axis has logarithmic scale.)} \label{fig6}
\end{minipage}
\hfill
\begin{minipage}[t]{0.45\linewidth}
\centering
\includegraphics[width=\textwidth]{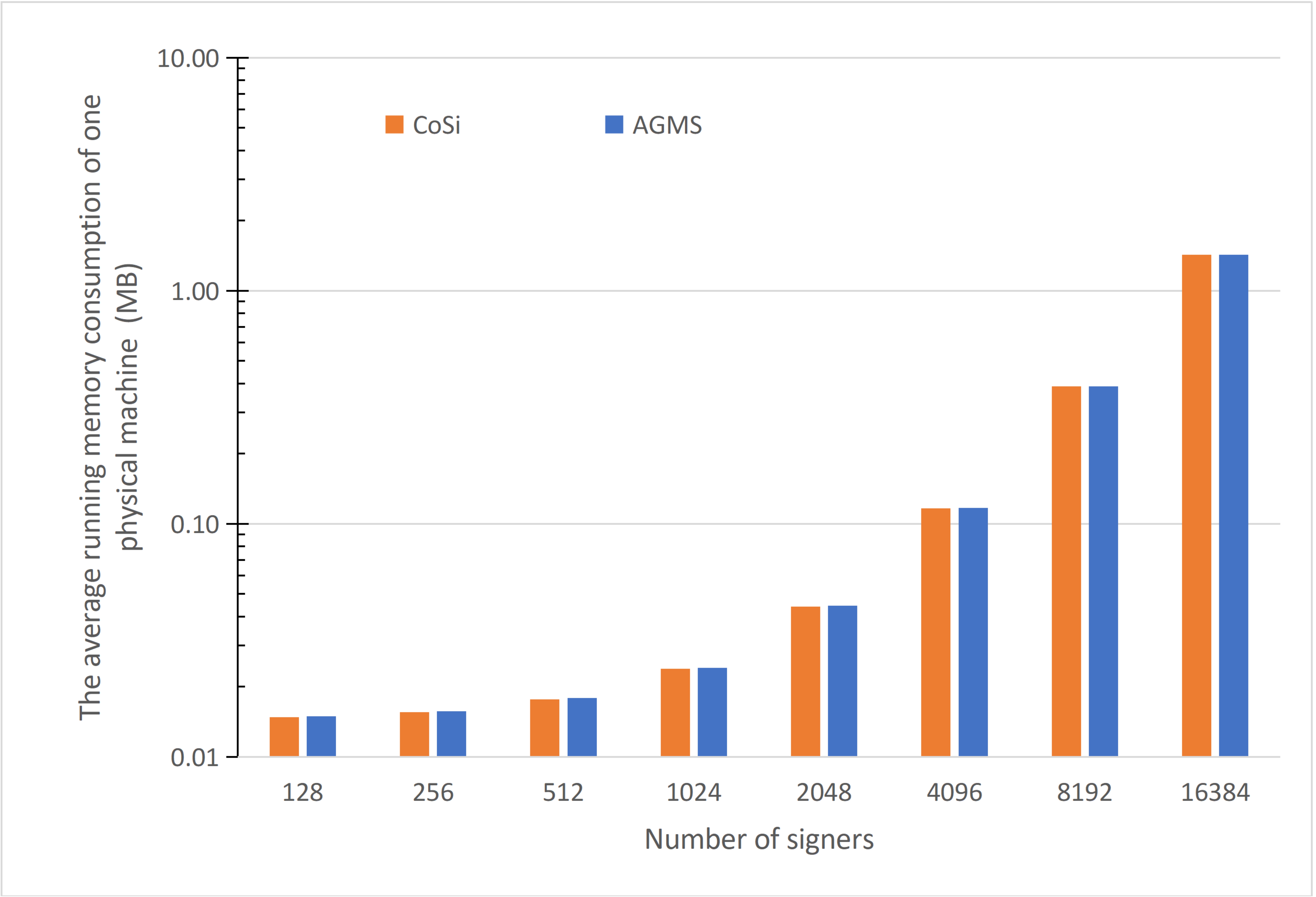}
\caption{Memory consumption of CoSi and AGMS ($y$-axis has logarithmic scale.)}
\label{figmemory}
\end{minipage}
\end{figure*}

Finally, as the leader has heavier computation load in signing algorithm than any other signers, we further test the computation time on a leader node of CoSi and the proposed AGMS in signing algorithm. In this experiment, we also divide signing algorithm into two phases: the former consists of the Announcement and Response phases, and the latter consists of Commitment and Challenge phases. The corresponding results are shown in Fig. \ref{fig5}. We clearly see that it takes much more time for the latter phases than the former phases, since the elliptic curve multiplication is much more complicated than scalar multiplication. Because the Commitment and Challenge phases can be precomputed in the proposed AGMS scheme, while CoSi needs to run all the phases in a sequential way, the proposed AGMS scheme runs absolutely faster than CoSi when we only focus on the computation time on a leader node in online signing phase. The total running time for the online signing phase of the two schemes are compared in Fig. \ref{fig6}.

Memory consumption is another factor to evaluate performance. On the group of 32 physical machines with the above configurations, we test the memory consumption in signing and verification algorithms of CoSi and AGMS with a total amount of signers ranging from 128 to 16384. From Fig. \ref{figmemory}, we can see that on one physical machine the memory consumption of CoSi and AGMS is very similar. Furthermore, as the vast majority of memory consumption is in offline signing phase, we have rather low memory consumption in online signing phase, which is very friendly to low-power devices.

\section{Application to Fabric}
Fabric \cite{11DBLP:conf/eurosys/AndroulakiBBCCC18} is a permissioned Blockchain platform, where a CA (Certificate Authority) is introduced to manage the members, and every node needs to make a request for membership to CA before it joins the network.
Digital signature algorithm ECDSA (Ellipse Curve Digital Signature Algorithm) is widely adopted in Fabric to guarantee the validity of transactions. To avoid inconsistency in transaction states, the client needs to collect enough number of signatures from different endorsers satisfying the endorsement policy in Fabric. If the endorsement policy requires a large number of endorsers, the number of signatures would be large, and the overhead of signature verification would be high. In this case, the current mechanism of Fabric will lead to significant drops of the transaction efficiency.

\begin{figure}
\includegraphics[width=0.5\textwidth]{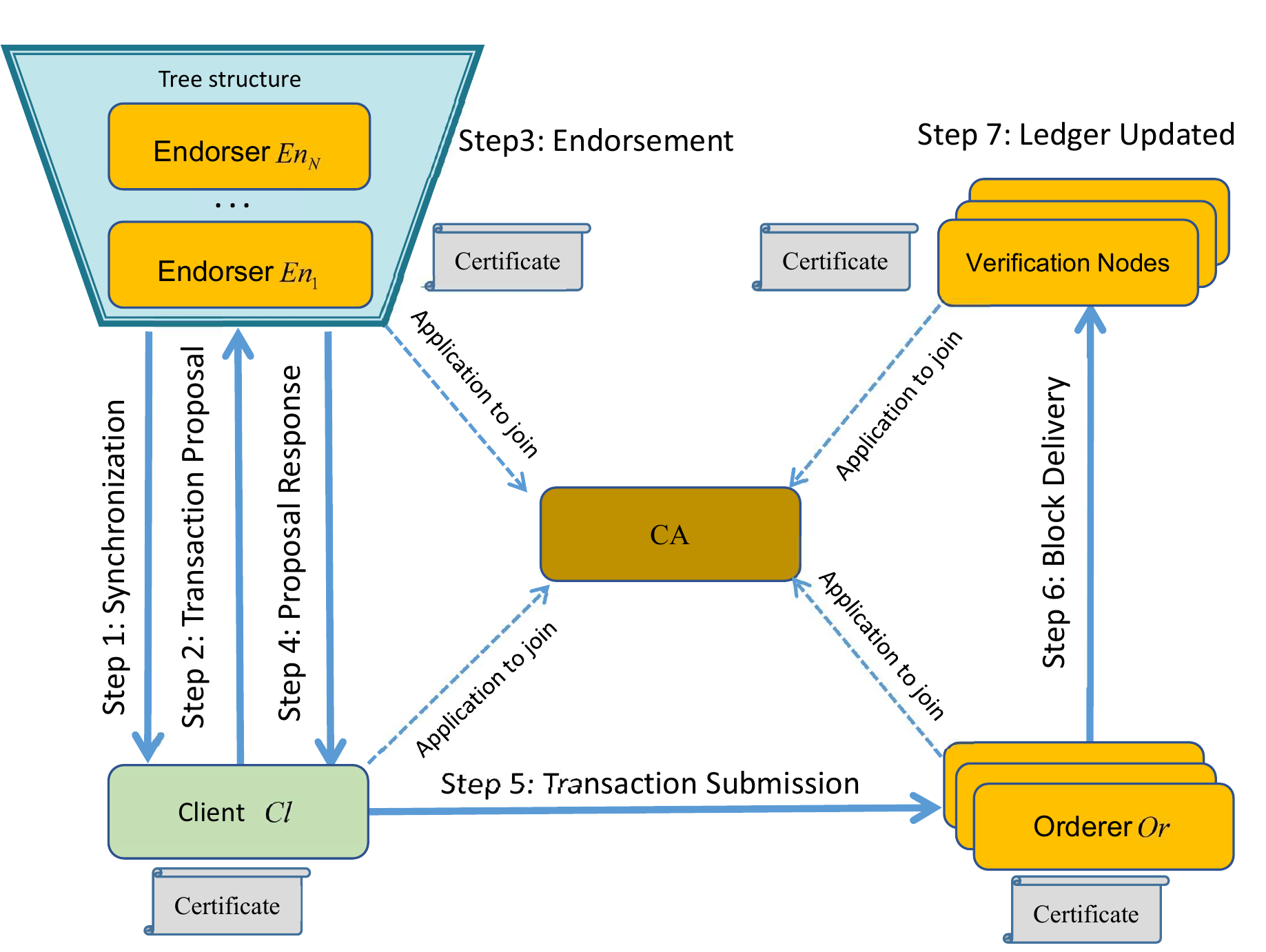}
\caption{The revised Fabric transaction process}
\label{figtransaction}
\end{figure}

Therefore, we try to introduce the proposed AGMS scheme into Fabric to optimize the current transaction process. In this paper, we implement the proposed AGMS on Fabric v1.0. In order to avoid confusion, we name original Fabric v1.0 as the default Fabric, and Fabric with AGMS as the revised Fabric. Compared to the default Fabric transaction process, we adopt our multi-signature scheme AGMS to replace ECDSA and add one synchronization step to run smoothly in the revised Fabric transaction process.
We assume the client as \(Cl\), the endorser as \(En_i\), and the orderer as \(Or\). We also define \(C_i\) as the set of children of one endorser \(En_i\),  \(P_i\) as the parent of the endorser \(En_i\), and \(N\) as the number of endorsers required by endorsement policy. As shown in Fig. \ref{figtransaction}, the revised Fabric transaction process can be described as follows.

Firstly, CA uses \(\textsf{Pg}(\kappa)\) to output \(par=(\mathbb{G},g_1,q)\). And then, each node uses \(\textsf{Kg}(par)\) to generate its own public/private key pair \((pk,sk)\). Before a node joins the Fabric network, CA additionally uses \(\textsf{KVf}(par,pk)\) to verify the validity of the node's identity and its public key. If the result is true, CA issues a certificate to the node so that it can successfully join the network. Otherwise, CA rejects the node, meaning that the node has no right to join the network of Fabric.

\noindent\textbf{Step 1: Synchronization.} All the endorsers \(En_i(i=1,\cdots,N)\) designated by endorsement policy can work as a sub-group in a spanning tree structure \(\tau\). They can synchronize the block information and implement phase 1 of \(\textsf{Sign}(par,(pk_i,sk_i),m,\tau)\). The client \(Cl\) works as the leader, implementing phase 2 of \(\textsf{Sign}(par,(pk_i,sk_i),m,\tau)\) to produce a common challenge \(c\), which acts as a part of the joint signature and is sent to each endorsers. The aggregated public key \(\tilde{X}\) is also computed in this section by \(\textsf{KAg}(\mathcal{PK})\).

\noindent\textbf{Step 2: Transaction Proposal.} When the client \(Cl\) needs to request a transaction \(m\), it firstly implements phase 3 of \(\textsf{Sign}(par,(pk_i,sk_i),m,\tau)\), sending the transaction proposal of \(m\) to the designated endorsers
\(En_i(i=1,\cdots,N)\) in a sub-group in a top-down way.

\noindent\textbf{Step 3: Endorsement.} When the endorser \(En_i\) receives a proposal from the client \(Cl\), it first uses \(\textsf{KVf}(par,pk)\) to check validity of the client \(Cl\)'s identity, then simulates the transaction implementation and signs the transaction proposal with its own private key and the previous common challenge \(c\). Finally, the endorser \(En_i\) implements phase 4 of \(\textsf{Sign}(par,(pk_i,sk_i),m,\tau)\), computing the partial response value \({s}_i\).

\noindent\textbf{Step 4: Proposal Response.} Then, all the designated endorsers \(En_i(i=1,\cdots,N)\) proceed to implement phase 4 of \(\textsf{Sign}(par,(pk_i,sk_i),m,\tau)\), sending back the proposal response bottom-up. The client \(Cl\) only needs to collect all the proposal responses from its children endorsers \(j\), which includes the simulated transaction results and the partial response values \(\tilde{s}_j\). When all the proposal responses are received, the client \(Cl\) checks the transaction results and computes \(S=\tilde{s}_{Cl}=s_{Cl}+\sum_{j\in C_{Cl}}\tilde{s}_j\). Finally, the client \(Cl\) successfully produces a joint signature \(\sigma=(c,S)\) representing the client \(Cl\) and all the designated endorsers \(En_i(i=1,\cdots,N)\). This joint signature can be easily verified by all nodes including the client \(Cl\) itself, so as to check whether it satisfies the endorsement policy.

\noindent\textbf{Step 5: Transaction Submission.} If the joint signature is valid, the client \(Cl\) sends the final transaction proposal and response to an orderer \(Or\).

\noindent\textbf{Step 6: Block Delivery.} The orderer \(Or\) orders the transactions from different clients into blocks and broadcasts them on the network;

\noindent\textbf{Step 7: Ledger Updated.} All the nodes on the network need to use \(\textsf{Vf}(par,\tilde{X},m,\sigma)\) to verify the block information and update synchronously.



\begin{figure*}
\begin{minipage}[t]{0.45\linewidth}
\centering
\includegraphics[width=\textwidth]{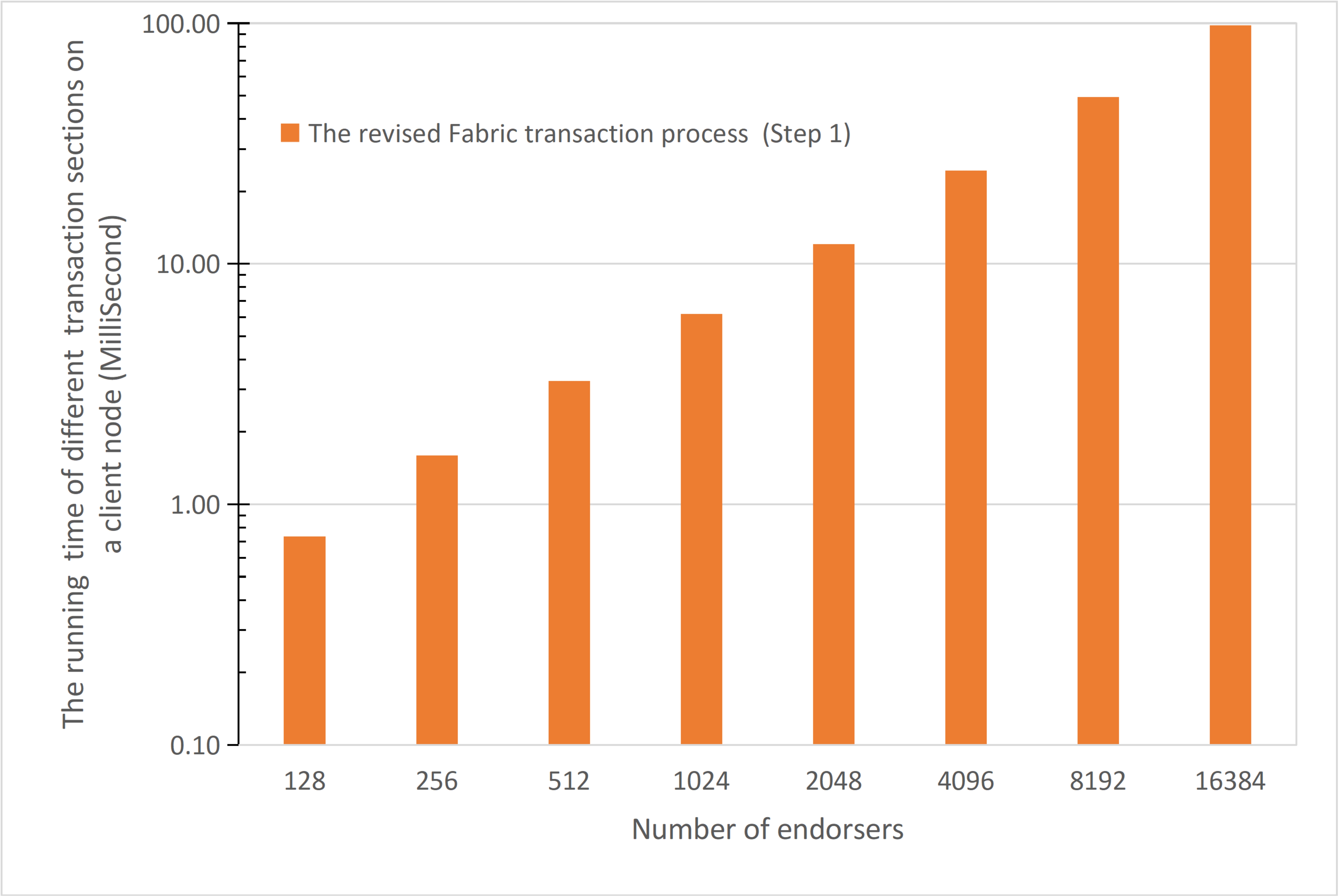}
\caption{The CPU running time of Step 1 on a client node in the revised Fabric transaction process ($y$-axis has logarithmic scale.)} \label{figHY1}
\end{minipage}
\hfill
\begin{minipage}[t]{0.45\linewidth}
\centering
\includegraphics[width=\textwidth]{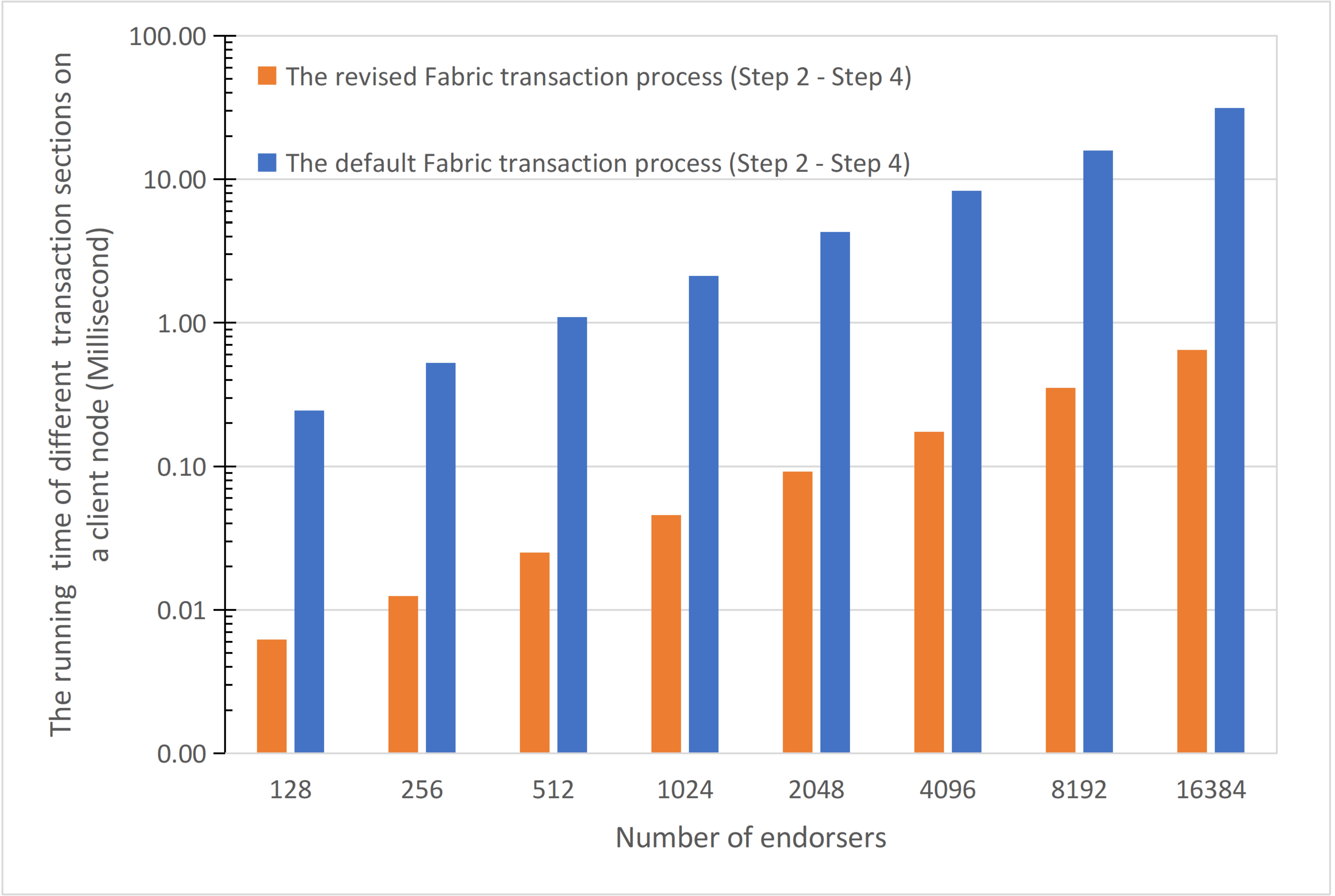}
\caption{The CPU running time from Step 2 to Step 4 on a client node between the default Fabric transaction process and the revised Fabric transaction process ($y$-axis has logarithmic scale.)} \label{figHY2}
\end{minipage}
\end{figure*}

\begin{figure*}
\begin{minipage}[t]{0.45\linewidth}
\centering
\includegraphics[width=\textwidth]{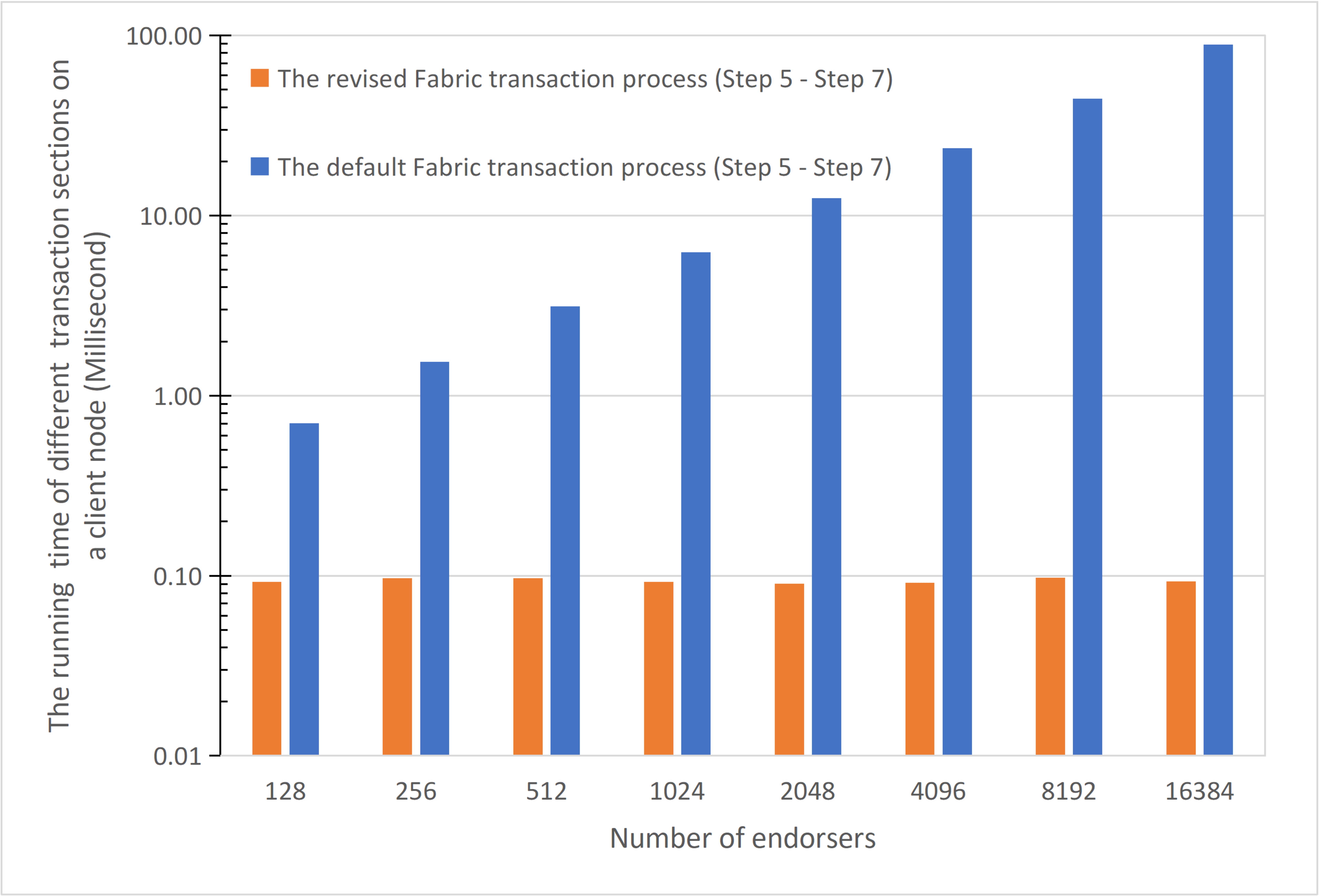}
\caption{The CPU running time from Step 5 to Step 7 on a client node between the default Fabric transaction process and the revised Fabric transaction process ($y$-axis has logarithmic scale.)} \label{figHY3}
\end{minipage}
\hfill
\begin{minipage}[t]{0.45\linewidth}
\centering
\includegraphics[width=\textwidth]{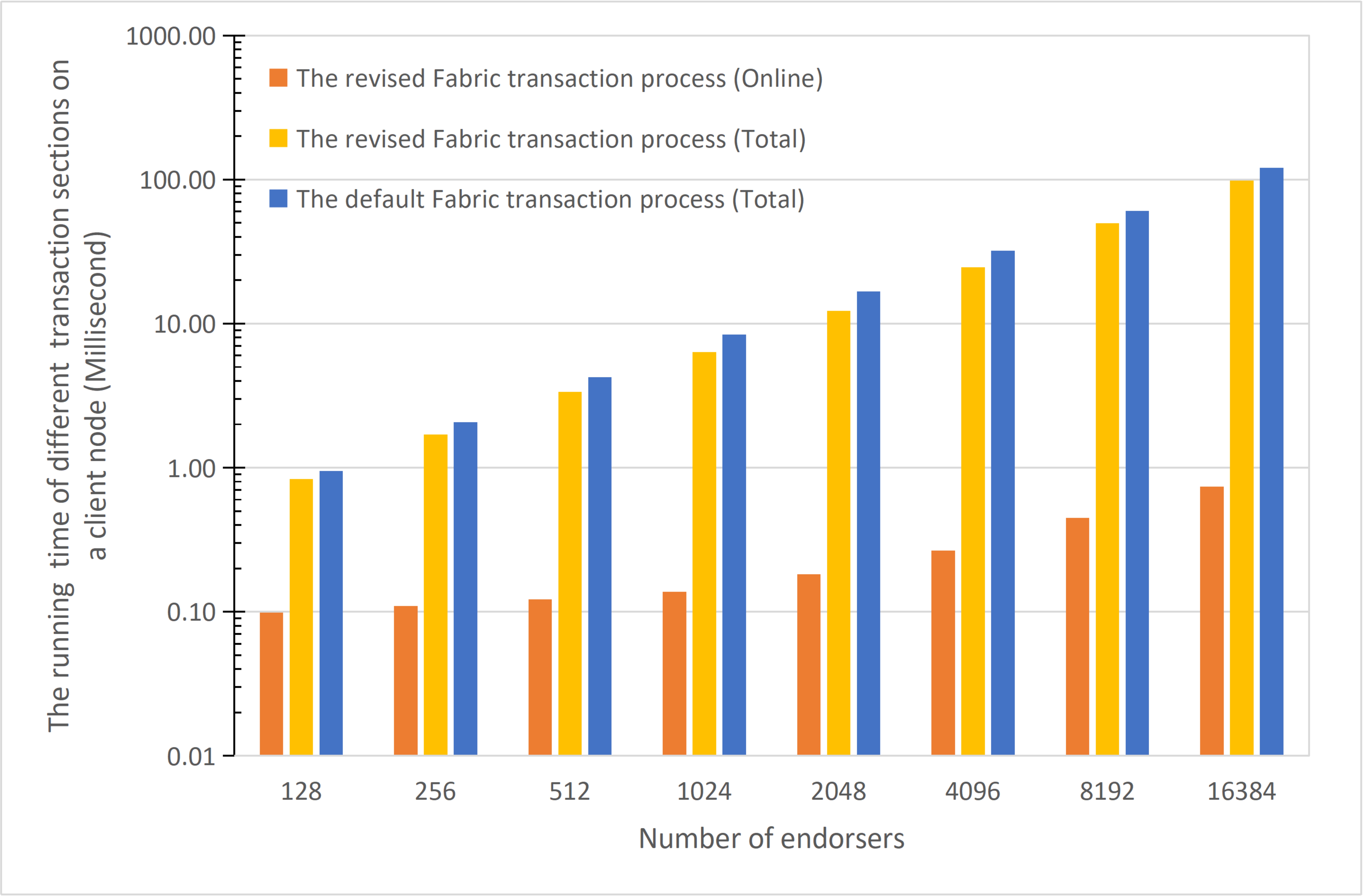}
\caption{The total CPU running time on a client node between the default Fabric transaction process and the revised Fabric transaction process ($y$-axis has logarithmic scale.)} \label{figHY4}
\end{minipage}
\end{figure*}

Some relevant experiments are shown in Fig. \ref{figHY1}, Fig. \ref{figHY2}, Fig. \ref{figHY3} and Fig. \ref{figHY4} respectively. We mainly test the running time of signing algorithm in different transaction sections on a client node for the default Fabric and the revised Fabric. All the configurations are the same as those in Section VI. We assume that we can set different numbers of endorsers without limitation and there is no delay in communication. Fig. \ref{figHY2} shows that, compared to the default Fabric, the revised Fabric transaction process runs much faster when a transaction comes. This is because the revised transaction process runs Step 1 shown in Fig. \ref{figHY1} in advance, which does not exist in the default Fabric. This step leads to a much faster online signing.
From Fig. \ref{figHY3}, in terms of Step 5 to Step 7, as the verification algorithm of ECDSA is implemented one time for each endorser, the CPU running time of the default Fabric increases linearly with the number of endorsers. But in the revised Fabric, the verification algorithm is only implemented once regardless of the number of endorsers. Thus, the CPU running time is almost constant.
Therefore, we can take advantage of this extra time to implement Step 1, and the total time of the revised Fabric transaction process is still shorter than that of the default Fabric. The results are shown in Fig. \ref{figHY4}.
In general, by applying the proposed multi-signature scheme AGMS to replace ECDSA in the default Fabric, the revised Fabric transaction process has faster online signing and verification performance and smaller storage space, so that we can achieve the goal of improving the transaction efficiency and reducing the transaction storage in a block.

Please note that although the proposed multi-signature scheme, AGMS, is implemented on Fabric, a permission based Blockchain platform, AGMS is also useful for permissionless Blockchain platforms, as it's not based on the assumption of Trusted Authority (TA). Taking a public and permissionless Blockchain Bitcoin as an example, there exists Multisig address \cite{andresen2011m}, which is the hash of $n$ public keys $(pk_1,pk_2,\ldots,pk_n)$. To spend funds associated with this address, one creates a transaction containing signatures from these $n$ public keys $(pk_1,pk_2,\ldots,pk_n)$. Authors in \cite{DBLP:journals/dcc/MaxwellPSW19} use multi-signature to aggregate multiple signatures into a joint one, so as to shrink the size of transaction data associated with Bitcoin Multisig addresses. Compared to the permissioned application, without a CA verifying the nodes' identities and permitting the entrance to Blockchain, the probability of attacks would increase. Nevertheless, the attacks can still be identified by the key verification and signature verification algorithms, which is guaranteed by the security of the proposed multi-signature schemes.

\section{Conclusion}
This paper proposes two multi-signature schemes based on Gamma signature. Compared to CoSi, the most popular multi-signature scheme based on Schnorr signature, the proposed schemes achieves enhanced security, higher online efficiency and similar scalability. We also apply the proposed AGMS to improve the transaction process of Fabric, so that the efficiency and throughput of Fabric are enhanced.

Undoubtedly, there are some limitations for the proposed multi-signature schemes in real-life implementation. If there exists tamper or forge in the multi-signature, the joint signature cannot pass the verification algorithm. However, the nodes in the tree need to verify the partial responses top-down to find out the malicious signer, which would increase the running costs. If the multiple signers are chosen in rotations, the malicious singer continuously sending wrong responses would be identified efficiently, leading to negligible attack probability. In addition, the revised Fabric transaction process is only suitable for the case where the endorsement policy is set as ``AND'', but not for ``OR'', ``NOT''. These limitations will be investigated more in-depth in our future work.


%

\ifCLASSOPTIONcaptionsoff
  \newpage
\fi



\bibliographystyle{IEEEtran}
\bibliography{refer}
%

%
\begin{IEEEbiography}[{\includegraphics[width=1in,height=1.25in,clip,keepaspectratio]{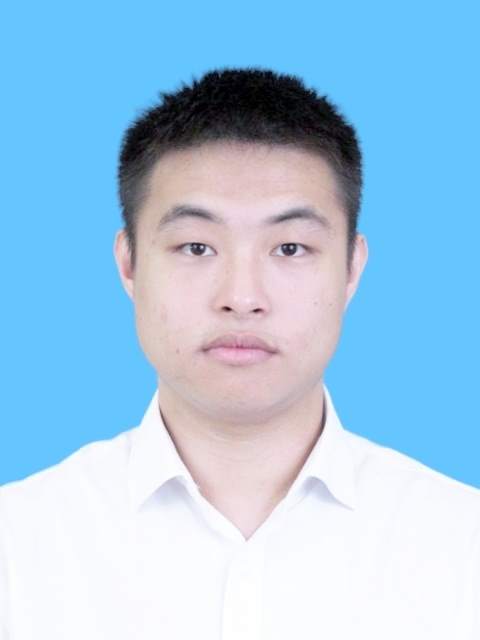}}]{Yue Xiao}
is a postgraduate student of College of Electronics and Information Engineering, Shenzhen University, China. He got the B.S. degree in telecommunication engineering from Guangdong Ocean University, China, in 2017, and the M.S. degree in information and telecommunication engineering from Shenzhen University, China, in 2020. His current research interests include cryptography technology and security in the Blockchain.
\end{IEEEbiography}

\begin{IEEEbiography}[{\includegraphics[width=1in,height=1.25in,clip,keepaspectratio]{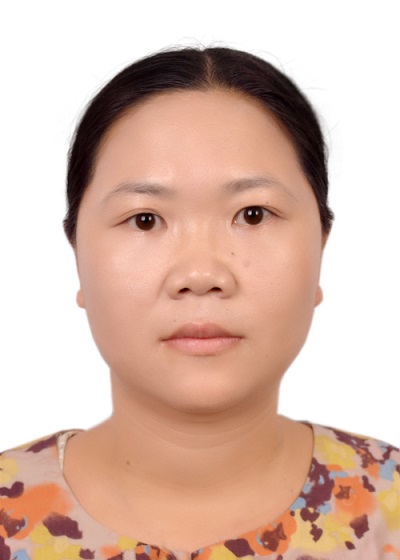}}]{Peng Zhang}
is an associate professor of College of Electronics and Information Engineering, Shenzhen University, China. She got the Ph.D. degree in signal and information processing from Shenzhen University, China in 2011. Her current research interests include cryptography technology and security in the Blockchain, Cloud Computing, IoT. She has published more than 30 academic journal and conference papers.
\end{IEEEbiography}

\begin{IEEEbiography}[{\includegraphics[width=1in,height=1.25in,clip,keepaspectratio]{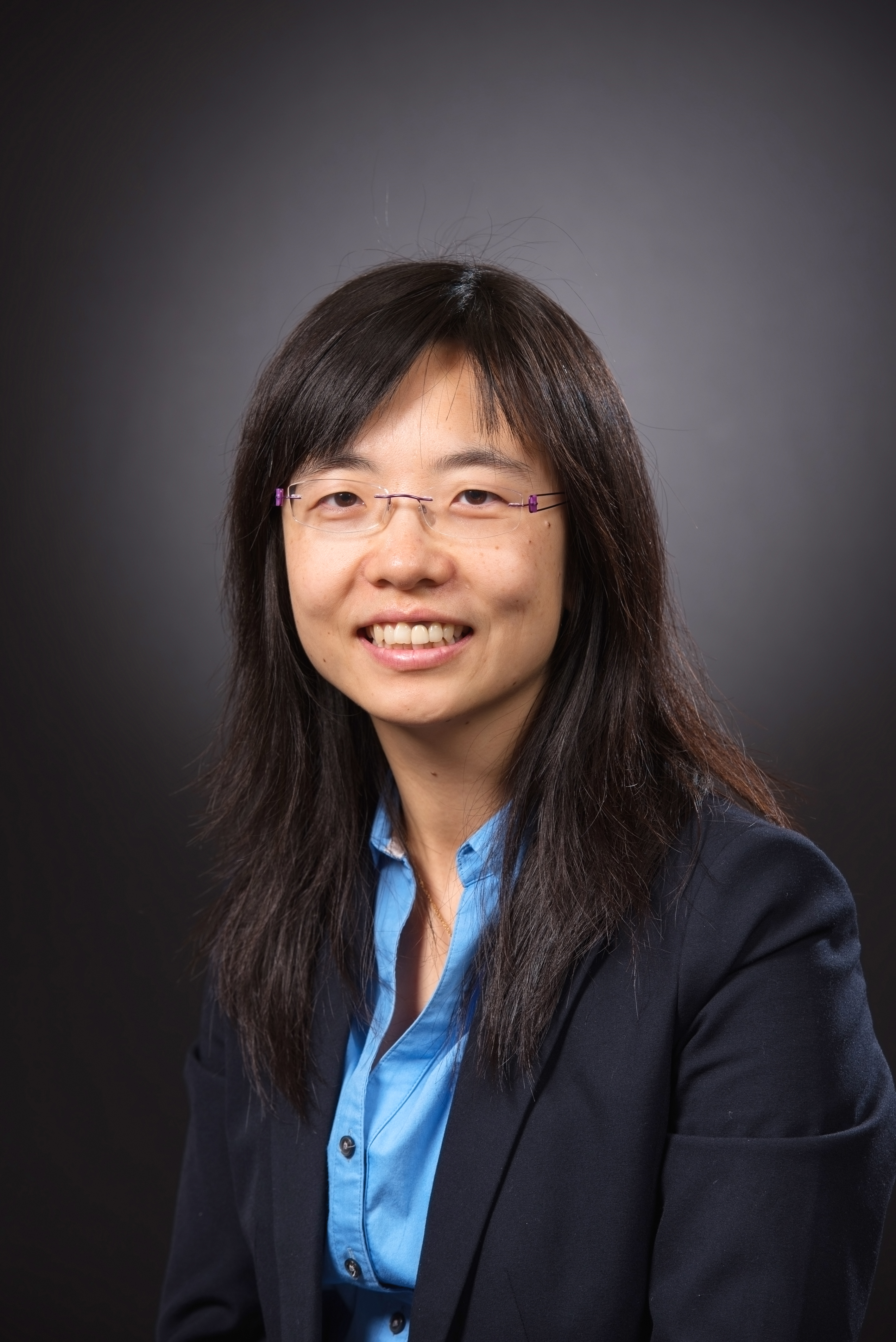}}]{Yuhong Liu}
is an Associate Professor at Department of Computer Engineering Santa Clara University. She received her B.S. and M.S. degree from Beijing University of Posts and Telecommunications in 2004 and 2007 respectively, and the Ph.D. degree from University of Rhode Island in 2012. Her research interests include trustworthy computing and cyber security of emerging applications, such as online social media, Internet-of-things, and Blockchain.
\end{IEEEbiography}
\end{document}